\documentclass[a4paper,
onecolumn,  unpublished, % accepted=yyyy-mm-dd,
superscriptaddress,11pt,noarxiv]{quantumarticle} 
\pdfoutput=1
\usepackage[utf8]{inputenc}
\usepackage[english]{babel}
\usepackage[T1]{fontenc}
\usepackage{amsmath}
\usepackage[colorlinks=true,urlcolor=blue,citecolor=blue,linkcolor=blue]{hyperref}
\usepackage{xcolor}
\usepackage{tikz}
\usepackage{accents}

\usepackage{enumitem}

\makeatletter % metadata for hyperref
 \hypersetup{pdftitle = {pdf title},
	     pdfauthor = {Frank Schmidt},
	     pdfsubject = {Quantum Communication},
	     pdfkeywords = {GKP, quantum cryptography, quantum repeater,  	     
				 PLOB, repeaterless bound,  
			    genuine, error-corrected, one-way, 
			    quantum entanglement, quantum internet,
			    error analysis,   
			    pure-loss channel, bosonic channel,
			    multimode, Fock, qudit, 
			    generalized Pauli, Weyl, Heisenberg,
			    quantum polynomial code, optimal distance, optimal spacing
			    }
	    }
\makeatother

\usepackage{cite}

\input{Qcircuit} % includes braket

\usepackage{subfig}
\usepackage[normalem]{ulem}   % crossed out: \sout
\usepackage{amsmath,amssymb}  % \lesssim
\usepackage{stmaryrd} % QECC brackets

\newcommand{\CSUM}{\textsc{{CSum}} }
\newcommand{\CPhase}{\textsc{{CPhase}} }

\newcommand{\unitspace}{\mskip3mu}

  \usetikzlibrary{shapes,arrows,spy,positioning,snakes}
\newcommand{\ut}[1]{\underaccent{\tilde}{#1}}
\renewcommand{\vec}[1]{\ut{#1}}  
  
\begin{document}
 
\title{Error-corrected quantum repeaters with GKP qudits}
\date{\today}		
\author{Frank Schmidt} \email{scfrank@uni-mainz.de}
\affiliation{Institute of Physics, Johannes Gutenberg-Universität Mainz, Staudingerweg 7, 55128 Mainz, Germany.}
\author{Daniel Miller} \orcid{0000-0003-2100-5612} 
\affiliation{Dahlem Center for Complex Quantum Systems and Institut f\"ur Theoretische Physik,
Freie Universit\"at Berlin,
Arnimallee 14, 14195 Berlin,
Germany.}

\author{Peter van Loock} \email{loock@uni-mainz.de}
\affiliation{Institute of Physics, Johannes Gutenberg-Universität Mainz, Staudingerweg 7, 55128 Mainz, Germany.}

\maketitle  

% ------------------------------------------------------------- -----------------  

\begin{abstract} 
The Gottesman-Kitaev-Preskill (GKP) code offers the possibility to encode higher-dimensional qudits into individual bosonic modes
with, for instance, photonic excitations.
Since photons enable the reliable transmission of quantum information over long distances and since GKP states subject to photon loss can be recovered to some extent, 
the GKP code has found recent applications in theoretical investigations of quantum communication protocols.
While previous studies have primarily focused on GKP qubits, the possible practical benefits of higher-dimensional GKP qudits are hitherto widely unexplored.
In this paper, we carry out performance analyses for three quantum repeater protocols based on GKP qudits
including concatenations with a multi-qudit quantum polynomial code. 
We find that the potential data transmission gains for qudits are often hampered by their decreased GKP error-correcting capabilities.
However, we also identify parameter regimes in which having access to an increased number of quantum levels per mode
can enhance the theoretically achievable secret-key rate of the quantum repeater. 
Some of our protocols share the attractive feature that local processing and complete error syndrome identification are realizable without online squeezing.
Provided a supply of suitable multi-mode GKP states is available,
this can be realized with a minimal set of passive linear optical operations,
even when the logical qudits are composed of many physical qudits.
\end{abstract} 
% ------------------------------------------------------------- -----------------  

\newpage
\tableofcontents

\newpage
\section{Introduction}
Quantum technologies rely on the availability of precisely controllable quantum systems, e.g., qubits, which can be realized with various physical implementations.
In 2000, Gottesman, Kitaev, and Preskill (GKP) proposed a method to encode finite-dimensional quantum systems (qudits) into quantum-mechanical harmonic oscillators~\cite{gkp}.
More recent theoretical developments include further proposals and assessments of GKP state preparation with superconducting devices~\cite{PhysRevX.11.011032, PRXQuantum.2.020101}.
After years of experimental progress, GKP qubits finally have been demonstrated in superconducting microwave cavities~\cite{Campagne-Ibarcq2020, eickbusch_fast_universal_2022, sivak_real_time_2022} and in the harmonic motion of ions~\cite{Fluehmann2019,deNeeve2022}.

In the optical domain, on the other hand, preparing GKP states is notoriously difficult.
The main problem is that reliable and strong nonlinearities are required but not readily available. 
In one approach, Gaussian Boson Sampling~\cite{opticalGKP1, opticalGKP2}, 
one exploits that measurements can induce nonlinear effects.
Here, Gaussian resource states are combined via passive linear optics and partially read out via photon-number resolving measurements. 
In this way, high-quality optical GKP states can be obtained, albeit only probabilistically. 
Gaussian Boson Sampling requires detectors with a sufficiently high level of photon-number resolution as well as increasingly complex linear circuits~\cite{opticalGKP1, opticalGKP2}. 
To shift the experimental burden associated with this, 
alternative approaches have been proposed~\cite{FukuiPRL2022, TakaseGaussBreeding2022}.
If non-Gaussian resource states or non-Gaussian optical elements are available, 
a recursive application of short linear circuits and homodyne measurements is sufficient for the preparation of GKP states~\cite{VasconcelosOL2010, WeigandPRA2018, BudingerCubic2022}.
There also exist alternatives which do not rely on measurements at all~\cite{HastrupNJPQPRA2021, BudingerCubic2022}.
A final option is to combine photon-subtraction- and homodyne-based elements to convert many-mode Gaussian cluster states into non-Gaussian few-mode states, which can be further processed into GKP states~\cite{EatonQuantum2022}.
Such an approach is compatible with measurement-based, continuous-variable quantum computation~\cite{BudingerCubic2022,PhysRevLett.97.110501}.

While the best method for creating optical GKP states has not yet been identified, it is safe to assume that their physical realization will require extremely sophisticated experimental procedures.
Once such technology is available, however, 
it will be comparatively straightforward to extend it to higher-dimensional GKP qudits and to concatenated multi-qubit or -qudit GKP codes.
For example, multiple GKP qubits can be entangled via Gaussian operations~\cite{gkp}.
Furthermore, ordinary beam splitters enable the generation of certain collective GKP ancilla states such as  Bell states with GKP qubits~\cite{WalshePRA2020} or qudits~\cite{gkp_syndrome}, as well as the collective detection of their 
error syndromes~\cite{gkp_syndrome}.
To guide such future experiments, we find it meaningful to investigate the performance of advanced multi-qudit GKP protocols in the realm of quantum communication.

The GKP encoding enables the correction of small displacement errors of the oscillator's quadratures, in particular, those that originate from typical Gaussian error channels such as amplitude damping or photon loss. 
However, large displacement errors cannot be avoided completely, especially for realistic, finitely-squeezed GKP states. 
This can cause misidentification of error syndromes, which leads to discrete logical errors on the affected GKP qudits.

In order to correct such errors, a higher-level quantum error-correcting code (QECC) can be employed to encode a few logical qudits into a larger number of physical GKP qudits~\cite{PRXQuantum.2.020101, gkp_analoginfo, FT_surface_gkp, toric-gkp, PhysRevX.8.021054, biased_gkp, GKP_LDPC}.
Hereby, the error correction capability of the higher-level QECC can benefit from analog information in the single-qudit GKP syndrome measurements~\cite{gkp_analoginfo, FT_surface_gkp, toric-gkp, PhysRevX.8.021054}. 
In order to satisfy the quantum singleton bound $n-k\geq2(d-1)$, every QECC with code parameters $\llbracket n,k,d \rrbracket$ must trade off the number of correctable (arbitrary) single-qudit errors against the number of physical qudits per logical qudit, which are given by $\lfloor(d-1)/2\rfloor$ and $n/k$, respectively~\cite{QMDS, PhysRevA.55.900, Calderbank1998}. 
An optimal trade-off is obtained by those QECCs that meet the quantum singleton bound with equality and are called maximum distance separable (MDS) codes.
While, for qubits, the only~\cite{qubit_QMDS} nontrivial (i.e., $d\geq3$ and $k\geq1$) MDS code encodes one logical qubit into five physical qubits~\cite{PhysRevLett.77.198},
there is a plethora of MDS codes for higher-dimensional qudits.
Such QECCs are explicitly available in the form of {quantum polynomial codes}, which exist for every qudit dimension being a prime power~\cite{PhysRevLett.83.648, Aharonov2008, Ketkar2006, CrossPhd}.

Currently, experimental realizations of long-distance quantum communication protocols 
are limited by the rapid decay of photonic signals that are sent through optical fibers.
This process is formally described by a pure-loss bosonic channel $\mathcal{L}(\eta)$, which arises from mixing the bosonic signal mode with an environmental mode in the vacuum state using a beam splitter with transmittance $\eta$.
In the long-distance limit of $\eta \rightarrow 0$, 
the secret-key capacity of the single-mode pure-loss channel scales linearly with $\eta$~\cite{TGW},
more precisely,
it is given by $-\log_2(1-\eta) \approx  1.44 \,\eta$~\cite{PLOB}.
In consequence, the secret-key rate of point-to-point quantum key distribution (QKD) is exponentially suppressed in the length $L$ of an optical fiber, which typically has a transmittance of  $\eta = \exp(-L/22\unitspace\text{km})$.

To overcome this problem,
quantum repeaters have been proposed~\cite{quantumrepeater_duerer}.
By introducing repeater stations, a long channel is split into multiple shorter ones.
To cope with the loss, 
different strategies have been conceptualized and, subsequently, been classified into three so-called generations of quantum repeaters~\cite{repeater_generations}.
These generations fundamentally differ in their speed of operation 
and in the level of technological maturity required for their realization.

\emph{First-generation quantum repeaters} are based on heralded, probabilistic entanglement distribution~\cite{quantumrepeater_duerer}. 
Once a Bell pair is successfully distributed between two neighboring repeater stations, it is stored in local quantum memories, where it resides until a second Bell pair, which connects the two repeater stations to a third one, is created.
Whenever two parts of different Bell pairs are present in a single repeater station, entanglement swapping can be executed, which results in a single Bell pair 
ranging over a larger distance.
This process is repeated until a long-distance Bell pair is shared between Alice and Bob.
In addition to channel loss, unavoidable operational gate and storage errors pose a challenge for quantum repeaters. 
To cope with such errors, first-generation quantum repeaters employ nested entanglement purification \cite{entanglement_purification}, a probabilistic protocol for the distillation of multiple low-fidelity Bell pairs into a smaller number of states with higher fidelities,
involving two-way classical communication.
In the worst case, entanglement purification has to be performed across the total distance $L$ of the entire quantum repeater chain, which slows down the achievable repetition rate to $c/L$ or less, where $c = 2.14\times 10^{8}\unitspace\tfrac{\text{m}}{\text{s}}$ is the speed of photons in fiber (for both classical and quantum signaling).

To avoid this slow-down, \emph{second-generation quantum repeaters} \cite{repeater_with_encoding} replace entanglement purification by QECCs for the local memories.
With this modification in place, the rate bottleneck is now posed by classical communication between neighboring repeater stations, which are separated by a distance of $L_0$.
Only after a failed entanglement distribution attempt has been heralded, the quantum memories can be freed up for the next attempt.
Therefore, the improved upper bound on the repetition rate is now given by $c/L_0$, which is typically on the order of $ 1 \unitspace\text{kHz}$ for $L_0\sim 100\unitspace\text{km}$  to
$1\unitspace\text{MHz}$ for $L_0 \sim 100\unitspace\text{m}$.
The only possibility to speed up the classical two-way communication
is to reduce $L_0$, i.e., to invest in a larger number of, realistically imperfect, faulty repeater stations whose quantum information must be consistently protected by the QECC.

Finally, \emph{third-generation quantum repeaters} enable ultrafast quantum communication as they dispense with the temporary storage of quantum information and classical two-way communication altogether~\cite{PhysRevLett.104.180503, PhysRevLett.112.250501, PhysRevA.89.032335}.
Instead, these repeaters employ QECCs to correct both channel losses and operational errors.
The repetition rates in this case are only limited by the speed of state preparations, local gate operations, and measurements in the individual repeater stations.
Whereas the preparation of QECC-encoded multi-photon states typically relies on some form of light-matter interaction, all other components of a third-generation quantum repeater can, in principle, be realized in an all-optical fashion~\cite{fabian_repeater_prl, fabian_repeater_pra, logicalBMefficiencies, AzumaNC2015, LeePRA2019}.

In this paper, we theoretically analyze the performance of third-generation quantum repeaters based on optical GKP qudits.
Our investigation also includes cases where the GKP code is concatenated with a higher-level QECC.
Here, we focus on quantum polynomial codes that previously have been considered in combination with multi-mode and Fock-encoded qudits~\cite{Muralidharan_2017, PhysRevA.97.052316, MHKB18,  MHKB19}.
For GKP-encoded states, similar performance studies have only been carried out in the special case of qubits~\cite{PhysRevA.63.022309, roz2020quantum, fukui2020alloptical}.
Our work thus closes the gap between these two approaches to a certain extent as it offers a treatment of the remaining case of GKP qudits.
The consideration of qudits, which can transmit more quantum information  per channel use than qubits, in the context of GKP and third-generation quantum repeaters is particularly attractive due to the existence of hardware-efficient GKP-qudit operations and syndrome extraction routines based on linear-optical elements alone.
In this way, the only fundamental experimental challenge that remains is to provide a supply of suitable multi-mode GKP ancilla states, a problem that can be tackled independently.

This paper is structured as follows.
In Sec.~\ref{sec:2}, we describe the details of our study: 
we begin with introducing the repeater protocols under investigation in Secs.~\ref{sec:protocol} and~\ref{sec:exp_repeaters} 
and proceed with 
our noise model in Sec.~\ref{sec:noise}.
In Sec.~\ref{sec:3}, we present the secret-key rates obtainable with the different GKP qudit repeater protocols and discuss the influence of various experimental parameters.
Finally, in Sec.~\ref{sec:4}, we summarize our results and conclude with a recommendation of the most promising quantum repeater protocol based on GKP qudits as identified in this work.

\section{Setting}
\label{sec:2}
GKP codes encode a $D$-dimensional qudit within the Fock space $\mathcal{F}$ of a quantum mechanical harmonic oscillator~\cite{gkp}.
We denote the annihilation operator of the oscillator by $\hat{a}$ and its quadrature operators by $\hat{p}=\tfrac{\text{i}}{\sqrt{2}}(\hat{a}^\dagger-\hat{a})$ and $\hat{q}=\tfrac{1}{\sqrt{2}}(\hat{a}^\dagger+\hat{a})$.
For simplicity, we focus in this paper on the square GKP code, which is defined as the $D$-dimensional subspace of $\mathcal{F}$ that is invariant under the action of $S_X = \exp(-\text{i} \sqrt{2\pi D}\hat{p})$ and $S_Z = \exp(\text{i} \sqrt{2\pi D} \hat{q})$.
By repeated non-destructive measurements of the stabilizer operators $S_X$ and $S_Z$, followed by appropriate displacement operations (or, at least, through tracking of the corresponding generalized Pauli frame), one can enforce the state of the oscillator to (effectively) remain 
in the GKP code space.
Since the logical Pauli operators of the square GKP code are given by
${X}=\exp(-\text{i} \sqrt{\tfrac{2\pi}{D}}\hat{p})$ and  ${Z}=\exp(\text{i} \sqrt{\tfrac{2\pi}{D}}\hat{q})$,
it is thereby possible (in the idealizing limit of perfect GKP states) to correct arbitrary displacement errors that are smaller than $\sqrt{{\pi}/{2D}}$ in magnitude.
To implement two-qudit gates between GKP qudits, 
one can utilize common two-mode Gaussian gates.
For example, on the level of GKP qudits, the bosonic  $\CSUM$-gate, $\exp\left(-\text{i}\hat{q}_1\hat{p}_2\right)$, acts as a two-qudit controlled-$X$ gate, $CX = \sum_{k=0}^{D-1} \ket{k}\bra{k}_1 \otimes X^k_2$.
A similarly defined $CZ$-gate is implemented by means of a $\CPhase$-gate, $\exp\left(\text{i}\hat{q}_1\hat{q}_2\right)$.  

\subsection{Repeater protocols} \label{sec:protocol}

Since GKP qudits can be encoded into photons, which are the ideal carriers of "flying" quantum information propagating at maximal speed,  they have been envisioned in the context of quantum communication~\cite{gkp_capacity, roz2020quantum, fukui2020alloptical}.
In this paper, we investigate certain quantum communication protocols that only require qudit Clifford operations and generalized Pauli measurements \cite{qudit_clifford}, which can be simply realized with GKP qudits by means of Gaussian optics and homodyne detection, respectively. 
More precisely, we analyze and compare the performance of three third-generation quantum repeater protocols introduced in the following subsections.
For each protocol, the term ``qudit'' may either refer to a bare (physical) GKP qudit or to an ensemble of multiple GKP qudits encoding a single (logical) qudit using a higher-level QECC, in particular, in combination with Knill's error-correction-by-teleportation procedure~\cite{gkp_syndrome, Knill_teleportation}.
Even in the absence of a higher-level QECC, 
our protocols represent instances of error-corrected (third-generation) quantum repeaters, as the availability of GKP syndrome information facilitates the correction of displacement errors to a certain extent.

 \begin{figure} \centering
 	\subfloat[]{\includegraphics[width=0.7\linewidth]{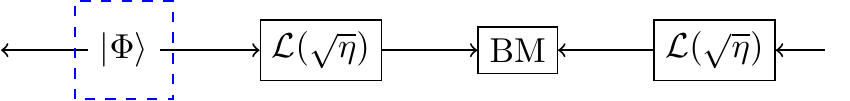}}\\
 	\vspace{5mm} 
 	
 	\subfloat[]{\includegraphics[width=0.7\linewidth]{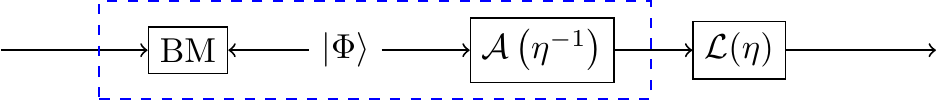}}\\
 	\vspace{5mm} 
 	
 	\subfloat[ ]{\includegraphics[width=0.8\linewidth]{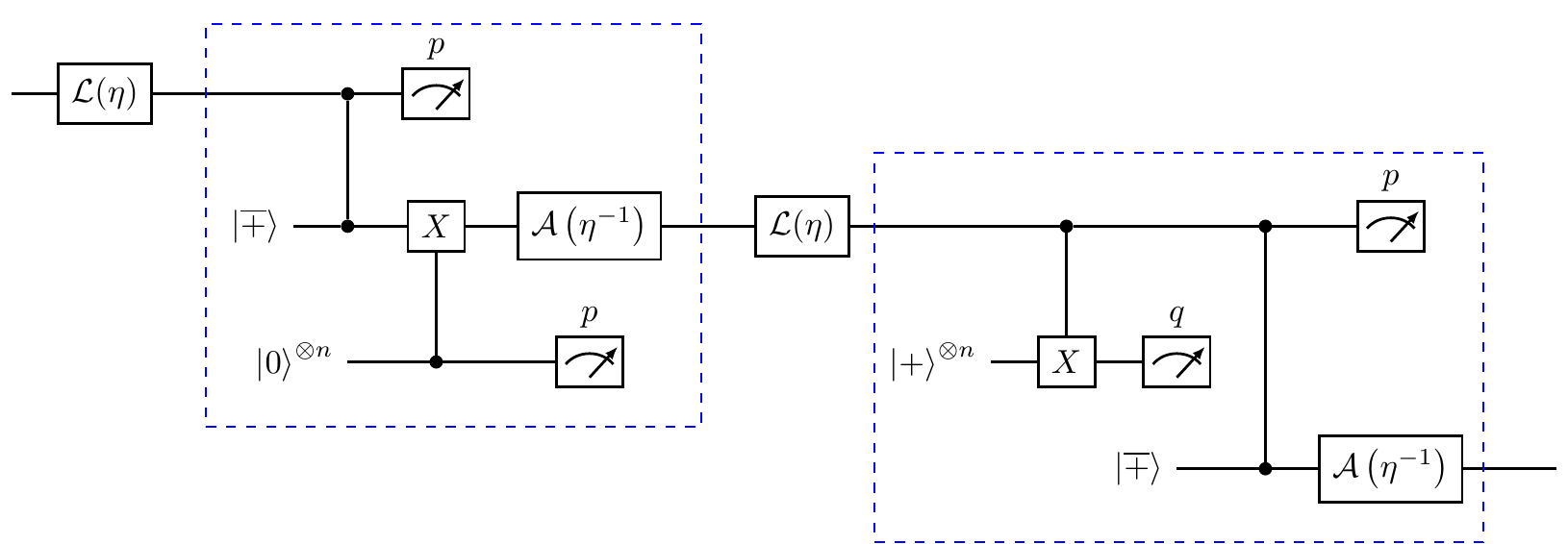}}
\caption{Unit cells of the quantum repeater protocols considered in this work.
The transmittance $\eta= \exp(-L_0/L_\text{att})$ of the bosonic pure-loss channel $\mathcal{L}(\eta)$ is exponentially suppressed in the distance $L_0$ between adjacent repeater stations (dashed blue boxes).
Here, the qudits can be either individual GKP qudits or logical qudits that are comprised of multiple GKP qudits by means of a higher-level $\llbracket n,1,d\rrbracket_D$ QECC.
\textbf{(a)} In the \textbf{two-way teleportation protocol}, 
every repeater station prepares two qudits in the maximally entangled state
$\ket{\Phi} = \tfrac{1}{\sqrt{D}}\sum_{k=0}^{D-1} \ket{k,k}$.
One of the two qudits is sent forward and the other one backward. %through the quantum repeater chain. 
After propagating a distance of $L_0/2$, at which each physical mode has been subject to a loss channel $\mathcal{L}(\sqrt{\eta})$,
a Bell measurement (BM) is performed.
\textbf{(b)} Also in the \textbf{one-way teleportation protocol}, two qudits are prepared in state $\ket{\Phi}$. 
In contrast to (a), only one of the qudits is sent to an adjacent repeater station.
To compensate for loss, a quantum-limited amplification channel 
$\mathcal{A}(\eta^{-1})$ with gain $\eta^{-1}$ is applied to each of the physical GKP modes.
After propagating a distance of $L_0$, a BM combines the forward-moving qudit with the stationary qudit of the subsequent repeater station.
\textbf{(c)} The \textbf{one-way half-teleportation protocol} is a GKP-adaptation of a previously-studied discrete-variable protocol~\cite{MHKB19}.
Here, we add measurements to convert displacement errors into Pauli errors.
Overlined ancilla states represent codewords of the higher-level $\llbracket n,1,d \rrbracket_D$ QECC, 
while ancilla states without overscore stand for GKP codewords. 
The $\overline{CX}$-gates correspond to transversal $\CSUM$-gates and the $\overline{CZ}$-gates corresponds to semi-transversal $\CPhase$-gates.
Measurements of $q$ and $p$ denote the measurement of the position and momentum quadrature, respectively. Loss and amplifier channels are again to be understood to act individually and independently on the physical GKP modes.}
 	\label{fig:repeaters}
 \end{figure}

\subsubsection{Two-way teleportation protocol with classical post-amplification}
\label{sec:2way_postamp}

The first of the three quantum repeater chains under investigation is portrayed in Fig.~\ref{fig:repeaters}~(a).
For this protocol, every repeater station prepares a pair of qudits in a (logical) Bell state.
One of the qudits is sent in the direction of the next repeater station, while the other one is sent backward.
In the middle between two neighboring repeater stations, the forward- and backward-propagating qudits are joined in a (logical) Bell measurement, which is implementable on the physical level with (transversal) beam splitters and two homodyne detectors per physical Bell measurement~\cite{gkp_syndrome}.
During the transmission from the repeater stations to the central Bell measurement apparatus,
the states of the qudits are altered due to the finite transmittance of the optical fiber channel. For the general case of many physical qudits representing one logical qudit, the optical loss channels act individually and independently (i.i.d.) upon the different modes of the physical multi-mode state that propagates through each fiber segment. 
To facilitate a direct comparison with the other protocols, we denote the channel transmittance by $\sqrt{\eta}= \exp(-L_0/2L_\text{att})$, as the relevant length of the fiber is given by $L_0/2$ here.
Throughout this paper, $L_0$ denotes the distance between two adjacent repeater stations,
and $L_\text{att}=22\unitspace\text{km}$ is the attenuation length of a typical fiber at the telecommunication wavelength of $1550\unitspace\text{nm}$.
In order to compensate for the loss-induced state change (with damped quadrature amplitudes), the classical measurement signal of the Bell measurements needs to be correspondingly amplified by a factor of $\sqrt{\eta}^{-1}$ before decoding the GKP syndrome.
Overall, this protocol produces an imperfect Bell pair ranging from one end of the repeater chain to the other.
Note that classical communication is only needed for post-processing and, 
therefore, it does not slow down the repetition rates of this protocol. 
Further note that for the case of a logical qudit composed of many physical qudits, classical post-amplification is performed individually for each physical Bell measurement to obtain the syndrome of the higher-level QECC~\cite{gkp_syndrome}.

\subsubsection{One-way teleportation protocol with optical pre-amplification} 
\label{sec:1way_preamp}

As a modification of the protocol from Sec.~\ref{sec:2way_postamp}, we also consider a quantum repeater chain where the Bell measurements are executed within the repeater stations, see Fig.~\ref{fig:repeaters}~(b).
Here, only one qudit per Bell pair is transmitted through the fiber channel.
This time, the transmittance is given by $\eta= \exp(-L_0/L_\text{att})$ because the traveling distance of the photons now covers a full repeater segment, i.e., twice the distance as in the previous scenario.
To cope with the fiber losses, an optical pre-amplification channel $\mathcal{A}\left(\eta^{-1}\right)$
is i.i.d.~applied to each (physical) GKP mode before it is sent through the fiber; this step replaces the classical post-amplification of the measurement signal from Sec.~\ref{sec:2way_postamp}.

\subsubsection{One-way half-teleportation protocol with optical pre-amplification}
\label{sec:1way_preamp_half}

The utilization of a Bell measurement (protocols described in Sec.~\ref{sec:2way_postamp} and Sec.~\ref{sec:1way_preamp}) provides GKP syndrome information for both quadratures.
This facilitates the correction of displacement errors on the level of the (physical and logical) GKP qudits.
For the final repeater chain under consideration, 
on the other hand,
every repeater station is responsible for preparing and measuring only a single logical GKP qudit, see Fig.~\ref{fig:repeaters}~(c).
This protocol has two core components.
First, a lower-level GKP error correction converts naturally occurring Gaussian displacement errors into Pauli errors on the physical qudits, see Sec.~\ref{sec:noise_conversion}.
Second, a higher-level QECC is utilized to cope with the resulting Pauli errors.
At the start of the repeater chain, Alice prepares two higher-level logical qudits in the state $\ket{\overline +} =   \sum_{k}  \vert{\overline{k}}\rangle/\sqrt{D}$ and entangles them with a logical $\overline{CZ}$-gate.
Since we restrict ourselves to quantum polynomial codes, the $\overline{CZ}$-gate admits a semi-transversal implementation with favorable error-spreading properties~\cite{Aharonov2008}.
Alice stores one of the logical qudits and to the second one, she applies a quantum-limited amplifier with gain $\eta^{-1}$ to each of the physical GKP modes before she sends them jointly through a lossy fiber of transmittance $\eta$ to the first repeater station, where the incoming logical qudit is entangled with a new logical qudit in state $\ket{\overline{+}}$.
A subsequent destructive, (physical) quditwise $p$-measurement
effectively transfers the encoded quantum information onto the next qudit and simultaneously delivers syndrome information involving $X$-stabilizers.
These steps are then repeated at every repeater station.
Besides yielding higher-level $X$-syndromes, the $p$-measurements are also responsible for providing lower-level GKP syndrome information $p \mod \sqrt{{2\pi}/{D}}$.
The physical $CZ$-gates propagate Gaussian $p$-errors on one mode into $q$-errors on the next one.
To prevent these $q$-errors from merging with $q$-errors that occur at the subsequent transmission, we introduce an additional ancilla-based GKP syndrome measurement in every repeater station.
This can be done in multiple ways, as discussed in App.~\ref{app:half_teleportation_gkp_syndrome_placement}.
To complete the protocol, all measurement results are communicated to Bob, who applies a suitable correction operator depending on the measurement outcomes~\cite{MHKB18}. 
Assuming $N$ is even and in the absence of errors, this protocol is equivalent to $N/2$ teleportation subroutines spread over $N+1$ different laboratories. 
For this reason, we refer to this protocol as \emph{half-teleportation}.

\subsection{Some comments on potential realizations of qudit repeaters}\label{sec:exp_repeaters}
To compensate for fiber loss, it is crucial to amplify the signal.
For the two protocols in Secs.~\ref{sec:2way_postamp} and~\ref{sec:1way_preamp}, one may opt between optical pre-amplification and classical post-amplification.
For the {half-teleportation} protocol in Sec.~\ref{sec:1way_preamp_half},
on the other hand, optical pre-amplification is the only option.
This is because the GKP qudits need to be correctly scaled, i.e., they need to be in the GKP code space up to a displacement,
before the $CZ$-gate is applied.
Since classical post-amplification can be carried out conveniently in software, lacking this option may be considered as a disadvantage of the half-teleportation protocol.

While we analyze their performance for GKP qudits, these protocols can be straightforwardly adapted to other qudit encodings, such as multi-mode (MM) qudits, 
which have been experimentally demonstrated in the context of (repeaterless) higher-dimensional quantum key distribution in the form of 
orbital angular momentum~\cite{oam_experiment} and time-bin qudits~\cite{Zhong_2015}.
Two of the three repeater protocols under consideration rely on Bell measurements. For GKP qudits, a deterministic Bell measurement can easily be implemented with static linear optics by employing a balanced beam splitter and continuous-variable homodyne measurements. 
Experimental implementations of Bell measurements for MM-encoded qudits, on the other hand, are disproportionately more involved.
Moreover, deterministic $CX$-gates for MM qudits require strong nonlinearities that are typically mediated through auxiliary matter qudits, which reduces the achievable repetition rates to the order of MHz.
This is in stark contrast to all-optical implementations that can reach GHz repetition rates. 
An attempt to circumvent this shortcoming of MM qudits is based on probabilistic linear optical Bell measurements, enabling an all-optical error correction step at every repeater station~\cite{fabian_repeater_prl,fabian_repeater_pra,logicalBMefficiencies, AzumaNC2015,LeePRA2019}. 
Such probabilistic Bell measurements cannot exceed 50\% for MM qubits in the simplest setting without additional resources such as photonic ancilla states~\cite{Calsamiglia_2001,GriceBM,Ewert34BM}.
For a deterministic Bell measurement, nonlinear optics is required.
Furthermore, probabilistic unambiguous state discrimination measurement of the corresponding two-qudit Bell states, making only use of linear optics and photon counting without ancilla photons, is impossible for MM qudits with $D>2$~\cite{qudit_linear_optics1, qudit_linear_optics2}.
Therefore, overall, the GKP concept and the GKP-based QR protocols presented in this work represent a unique way to combine an increased communication capacity based on photonic qudit encoding with an enhanced loss (and error) robustness based on photonic qudit quantum error correction.

\subsection{Noise model}
\label{sec:noise}

GKP codes are designed to correct displacement errors.
As we review next, this allows us to model photon loss and imperfect GKP state preparation with incoherent Gaussian displacement channels.
For our error analyses, it will suffice to keep track of their variances. 

\subsubsection{Transmission loss and coupling inefficiencies}
\label{sec:loss_and_amplification}
 
The bosonic pure-loss channel $\mathcal{L}(\eta)$ is commonly used to model fiber loss and coupling inefficiencies in quantum communication protocols~\cite{RevModPhys.77.513, RevModPhys.84.621}.
When $\mathcal{L}(\eta)$ is applied to a GKP state, its quadratures are damped, which shrinks the GKP lattice.
To rescale the lattice, one has to amplify the signal.
Depending on whether this amplification is carried out optically before $\mathcal{L}(\eta)$, optically after $\mathcal{L}(\eta)$, or classically after the measurement of a quadrature operator, the effective error channel on the GKP subspace is altered.

For the one-way protocols in Secs.~\ref{sec:1way_preamp} and~\ref{sec:1way_preamp_half}, we
consider the usage of an optical amplification channel $\mathcal{A}(\eta^{-1})$.
If $\mathcal{A}(\eta^{-1})$ is applied \emph{after} $\mathcal{L}(\eta)$, 
the result is a Gaussian displacement channel with variance $\sigma^2 = (1-\eta)/\eta$~\cite{PhysRevA.63.022309}.
If $\mathcal{A}(\eta^{-1})$ is applied \emph{before} $\mathcal{L}(\eta)$, however, the variance is improved to $\sigma^2 = 1-\eta$, as this avoids amplifying noise that occurs during transmission~\cite{gkp_capacity}.
In our analyses of the one-way protocols, we will therefore consider the latter strategy.
Furthermore, we will assume a total transmittance of
$\eta_\text{tot}=\eta_\text{c}\exp\left(-{L_0}/{L_\text{att}}\right)$,
where $\eta_\text{c}$ denotes the efficiency for coupling into the fiber  ($\eta_\text{c}=0.99$ unless stated otherwise) and $L_\text{att}=22\unitspace\text{km}$ is the attenuation length.

For the two-way teleportation-based protocol in Sec.~\ref{sec:2way_postamp}, 
it is possible and beneficial to replace $\mathcal{A}(\sqrt{\eta}^{-1})$
with a classical amplification of the measured signal. 
Effectively, this turns the loss into a Gaussian error channel with variance
$\sigma^2={1/\sqrt{\eta_\text{tot}}} -1 $~\cite{fukui2020alloptical},
where $\eta_\text{tot}=\eta_\text{c}^2\exp\left(-{L_0}/{L_\text{att}}\right)$ takes into account that,
in a two-way protocol, two signals are coupled into the fiber.

\subsubsection{Approximate GKP state generation}
The second, important noise contribution arises during the preparation of GKP states.
In position basis, the state vector of an ideal square GKP qudit takes the form
\begin{equation}
    \ket{j}=\sum_{k\in\mathbb{Z}}\ket{\hat{q}=\sqrt{\frac{2\pi}{D}}\left(j+D  k\right)}\,,
\end{equation}
where $j\in\{0,\ldots, D-1\}$ labels a computational basis state.
These ideal states are unphysical as they are neither normalizable nor superpositions of finite-width peaks. To describe normalizable, physical instances of GKP states and eventually also predict real-world experimental performances, we instead consider approximate GKP states for
which multiple realizations have been proposed that are essentially\footnote{The state given in Eq.~\eqref{eq:GKP_def1} is not symmetric under exchange of position and momentum. 
However, this state can be squeezed by a factor of $\sqrt{1+\kappa^2\Delta^2}$ to obtain the parameterization given in Eq.~\eqref{eq:GKP_def2}.} equivalent~\cite{gkp, gkp_threshold_prl, approximimate_gkp}.
Normalizability can be restored using an overall slowly decaying Gaussian envelope and the delta peaks can be approximated  with (a still infinite number of) highly squeezed Gaussian peaks.
This results in approximate GKP states of the form
\begin{align}
      \ket{\Tilde{j}} \propto\sum_{k\in \mathbb{Z}}
      \exp\left({- \frac{\pi \kappa^2}{D}(j+Dk)^2}\right)
      \int_{-\infty}^\infty \text{d}q \, 
      \exp\left({-\frac{(q-\sqrt{\frac{2\pi}{D}}(j+Dk))^2}{2\Delta^2}}\right)
      \ket{\hat{q}=q},\label{eq:GKP_def1}
\end{align}
where $\Delta$ and $\kappa$ are squeezing parameters corresponding to the peaks' width in position and momentum representation, respectively.
Alternatively, $\vert{\Tilde{j}}\rangle$ can be interpreted as an ideal GKP state $\ket{j}$ to which coherent Gaussian displacements have been applied, i.e.,
\begin{align}
\ket{\Tilde{j}}
    \propto  \int_{\mathbb{R}^2} {\text{d}u\, \text{d}v} \,
    \exp\left(-\frac{1}{2}\left(\frac{u^2}{\gamma^2}+\frac{v^2}{\delta^2}\right) 
    + \text{i}\left(\frac{-u\hat{p}+v\hat{q}}{\sqrt{2}}\right) \right)
  \ket{j},\label{eq:GKP_def2}
\end{align} 
where the squeezing parameters $\gamma$ and $\delta$ are in one-to-one correspondence to $\Delta$ and $\kappa$, see Thrm.~1 in Ref.~\cite{approximimate_gkp}.
In this work, we only consider the symmetric case of $\gamma=\delta$.
As a further simplification, we assume incoherent Gaussian displacements with variance $\sigma_{\mathrm{sq}}^2$,
which can be understood as a twirling-approximation~\cite[App.~A]{FT_surface_gkp}. 
Numerical simulations confirm that such an approximation does not overestimate the approximate GKP state's fidelity~\cite{gkp_teleportation_twirling}.
Following Refs.~\cite{FT_surface_gkp,gkp_threshold_prl,kosuke_threshold}, we define the \emph{squeezing parameter} (given in dB),
\begin{align} \label{eq:squeezing_parameter}
     s_\text{GKP} = -10\log_{10}\left(\frac{\sigma^2_\text{sq}}{\sigma^2_\text{vac}}\right), 
\end{align} 
where $\sigma^2_\text{vac}=1/2$ denotes the quadrature variance of the vacuum state.

By means of a higher-level QECC, it is possible to concatenate multiple approximate GKP qudits, each of which is modeled by an ideal GKP state followed by Gaussian squeezing errors, into a single logical qudit.
The corresponding unitary encoding circuit may redistribute the error probabilities between the modes, which in principle leads to correlated errors~\cite{MHKB18}.
The resulting error probabilities have a complicated dependence on the selected encoding circuit, 
thus, they cannot be easily captured in full generality in our analytical model.
Therefore, we leave such details for future work.
For the purpose of the present investigation, we are satisfied with a noise model, where unphysical, ideal GKP states are first encoded using a higher-level QECC and, afterward, physicality is restored by 
applying Gaussian squeezing channels i.i.d.~to each qudit, as motivated above.

\subsubsection{Converting Gaussian noise into Pauli errors}
\label{sec:noise_conversion}

The purpose of the GKP error-correction step shown in Fig.~\ref{fig:repeaters}
is to discretize the continuous displacement errors that build up on the GKP qudits.
In general, a single-qudit Pauli error channel is completely described by its joint error probability distribution of $X$- and $Z$-errors~\cite{MHKB18}.
We denote such a distribution by
\begin{align}\label{eq:error_probability_distribution}
	\mathcal{P}(X,Z)=\begin{pmatrix}
	P(X^0,Z^0)&\dots  &P(X^0,Z^{D-1})  \\ 
	\vdots	& \ddots  & \vdots  \\ 
	P(X^{D-1},Z^0)& \dots & P(X^{D-1},Z^{D-1})
	\end{pmatrix} .
\end{align} 
Let us calculate, for a square-lattice GKP qudit, the Pauli error channel that results from a Gaussian noise channel with zero mean and a covariance matrix $\Sigma_\text{sq}=\sigma^2 \mathbb{I}$ (with respect to $q$ and $p$). 
We find that $X$- and $Z$-errors are independent because the same is true for the two Gaussian random variables describing $q$- and $p$-shifts.
In other words, the matrix $\mathcal{P}_\text{sq}(X, Z)=\mathcal{P}_\text{sq}(X)\otimes\mathcal{P}_\text{sq}(Z)$ factors into the  outer product of the error probability vectors that store the marginal distributions of $X$- and $Z$-errors. 
By symmetry of the square lattice, we have $\mathcal{P}_\text{sq}(X)=\mathcal{P}_\text{sq}(Z)$.
The probability to suffer $k\in\{0,\ldots, D-1\}$ shifts can be expressed as
\begin{align} \label{eq:pauli_shift_probability_gauss}
	P_\text{sq}(X^k,\sigma^2)&=\sum_{j\in\mathbb{Z}}\int_{\sqrt{\frac{2\pi}{D}}(jD+k-\frac{1}{2})}^{\sqrt{\frac{2\pi}{D}}(jD+k+\frac{1}{2})}\frac{1}{\sqrt{2\pi \sigma^2}}\exp\left(-\frac{q^2}{2\sigma^2}\right)dq\\
	&=\sum_{j\in\mathbb{Z}}\frac{1}{2}\left(\text{erf}\left(\sqrt{\frac{2\pi}{D}}\frac{jD+k+\frac{1}{2} }{\sigma}\right)-\text{erf}\left(\sqrt{\frac{2\pi}{D}}\frac{jD+k-\frac{1}{2} }{\sigma}\right)\right)\,,\nonumber
\end{align}
where $\text{erf}(x)=\frac{2}{\sqrt{\pi}}\int_{0}^x \exp(-q^2)dq$ is the error function.
For our purposes, it is sufficient to keep only the three terms with $|j|\le 1$.

\section{Secret-key rates of quantum repeaters}
\label{sec:3}
The central figure of merit that we employ to compare the performance of different repeater protocols is the secret-key rate (SKR) per channel use.
More precisely, we use $\log_2(D)-H(\mathcal{P})$, which is a lower bound on the two-way capacity~\cite{PLOB}, where $H(\mathcal{P})$ denotes the Shannon entropy of a Pauli error probability distribution $\mathcal{P}$ as in Eq.~\eqref{eq:error_probability_distribution}.
Note that this bound can be achieved by a qudit generalization (using $D+1$ bases, assuming $D$ to be prime) of the six-state protocol~\cite{PhysRevLett.81.3018} in the asymptotic limit, where almost every round the same basis is used~\cite{quditqkdprotocols}.  
Moreover, if $X$- and $Z$-errors are independent, the same rate is obtainable with a generalization of the BB84 protocol~\cite{bb84} (2 bases, arbitrary $D$).\footnote{The secret-key fraction is given by $I(A, B)-I(A, E)=\log_2(D)-H(\vec{q}_{01})-I(A, E)$, where expressions of the mutual information $I(A, E)$ between Alice and Eve are provided in Eqs.~(5) and~(7) of  Ref.~\cite{quditqkdprotocols}.}

\subsection{Repeater performance with GKP error correction only}
\label{sec:result_bare_gkp}

\begin{figure} \centering
	\subfloat[]{\label{fig:complete_L0=500m_Lmax=100000km_099coupling_lo_logx_preamp}\includegraphics[width=0.49\textwidth]{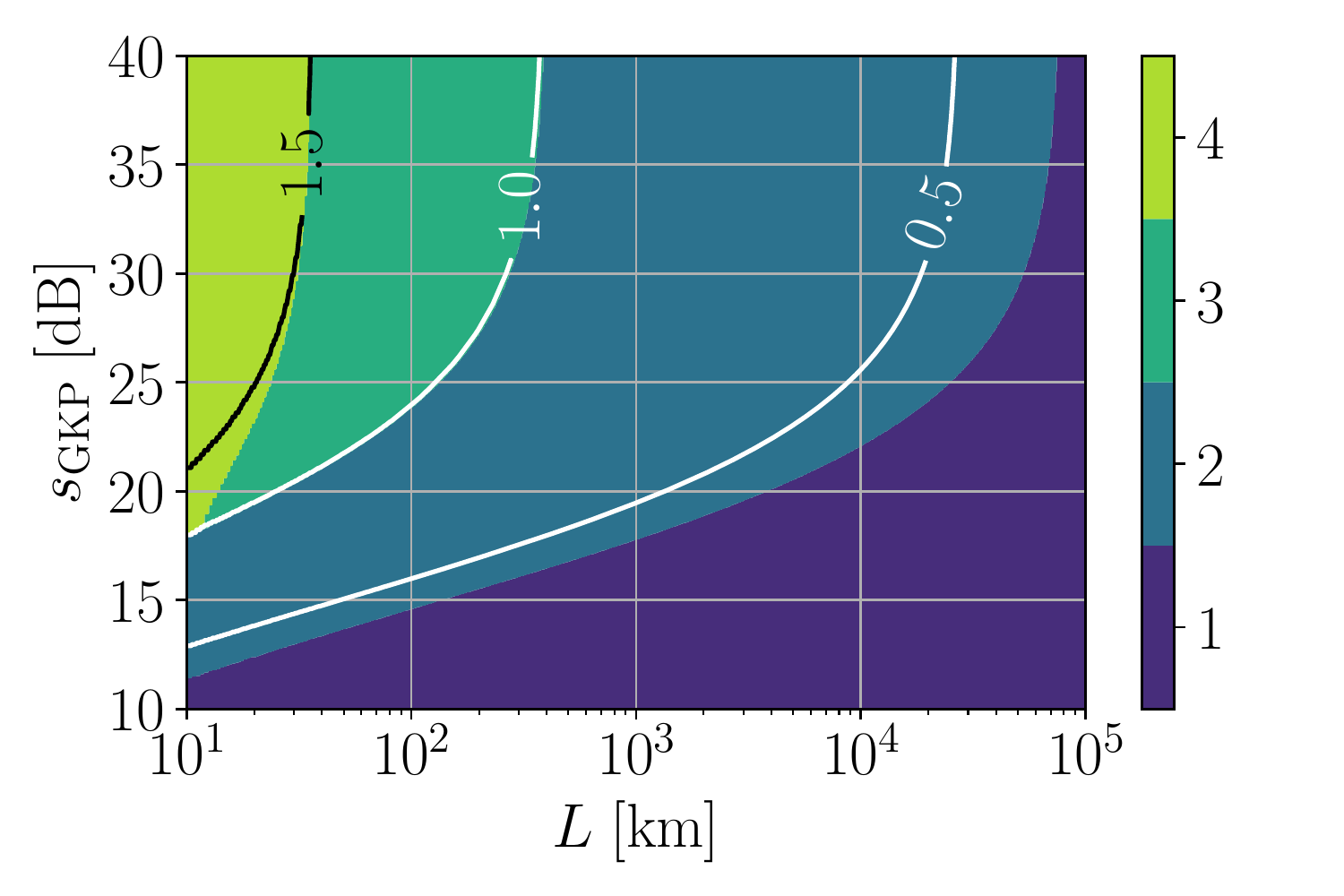}}
	\subfloat[]{\label{fig:complete_L0=500m_Lmax=100000km_099coupling_lo_logx_twowaycc_final}\includegraphics[width=0.49\textwidth]{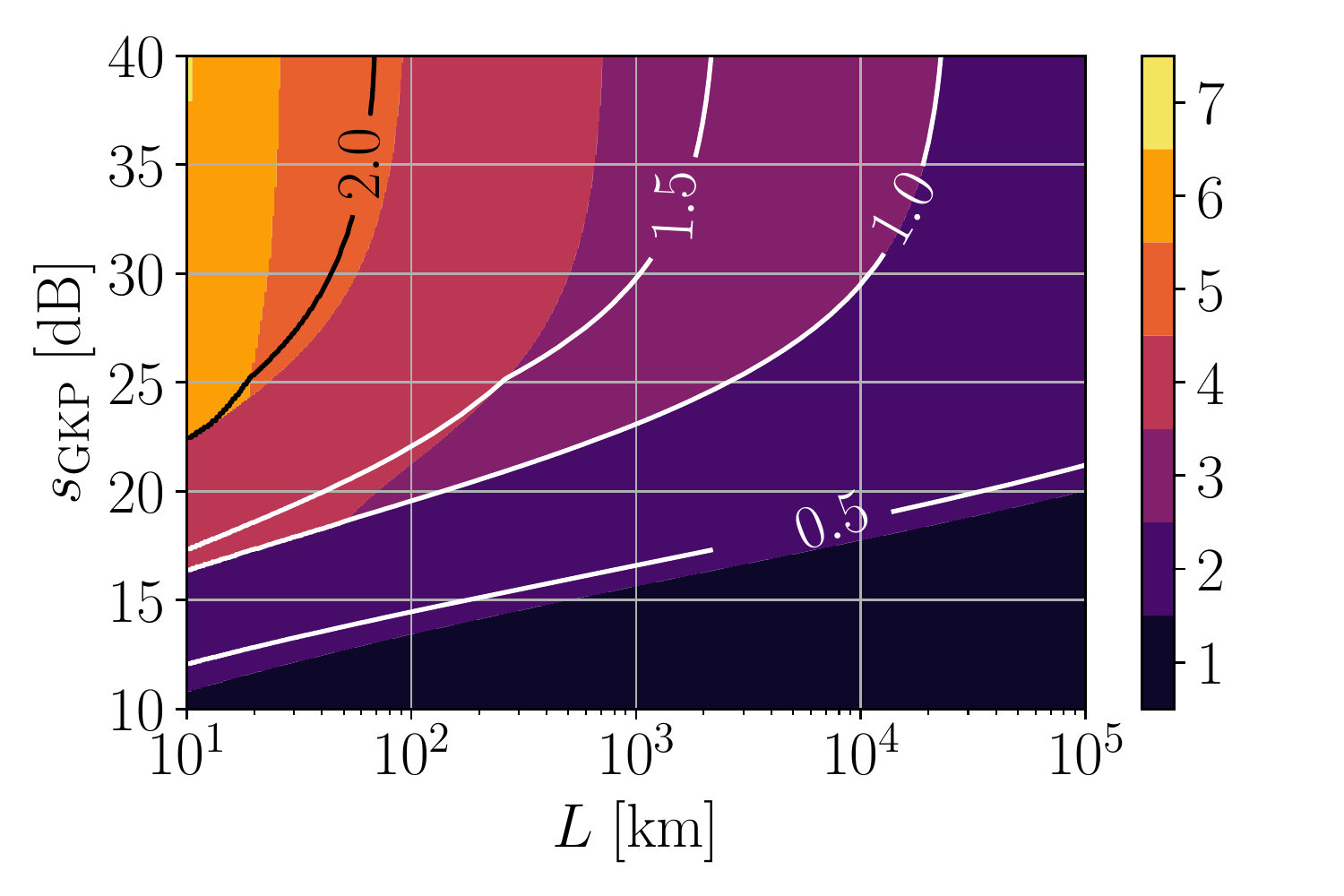}}
    \vspace{-35mm}
    \begin{minipage}{\textwidth} \small
        \hspace{68mm} $D$
        \hspace{69mm} $D$
    \end{minipage}
    \vspace{26mm}
	\caption{Optimal dimension $D$ of bare GKP qudits utilized in a quantum repeater line with coupling efficiencies $\eta_\text{c}=99\%$ and an intermediate repeater spacing of $L_0=500 \unitspace\text{m}$, where the (a)  one-way or (b) two-way teleportation protocol is used.
    For each choice of total repeater length $L$ and squeezing parameter $s_\text{GKP}$, 
    the qudit dimension is adjusted such that the SKR per channel use, $\log_2(D)-H(\mathcal{P})$, is optimized (inset lines). 
    In the parameter regions of $D=1$, it is not possible to generate secret keys.
}
	\label{fig:figa}
\end{figure}

For near-term applications, it is certainly more convenient to operate a quantum repeater with bare GKP qudits and not with multiple GKP qudits in a QECC.
To guide such initial experiments, we begin our discussion with this important special case.
For the two protocols considered with bare GKP qudits, which are described in Secs.~\ref{sec:2way_postamp} and~\ref{sec:1way_preamp}, 
lower-level error correction is performed via a teleportation step on the logical level of the GKP code, which leads to independent $X$- and $Z$-errors.
As mentioned above, the SKR per channel use is thus given by $\log_2(D)-H(\mathcal{P})$ not only for the generalized six-state protocol ($D$~prime) but also for the generalized BB84 protocol ($D$~arbitrary).
The precise value of $H(\mathcal{P})$  has a complicated dependence on the repeater spacing $L_0$, on the total repeater length $L$, on the squeezing parameter $s_\text{GKP}$ that characterizes approximate GKP states, and on the qudit dimension $D$.
However, we can numerically assess $H(\mathcal{P})$, see App.~\ref{app:error_analysis_bare}.
In Fig.~\ref{fig:figa}, we show the optimal choice (color-coded) of qudit dimension $D$ for different values of $L$ and $s_\text{GKP}$, 
where $L_0 = 500\unitspace\text{m}$ is fixed.
Using inset lines, we also display the corresponding (maximal) value of the SKR per channel use.
As expected, the key rate vanishes if the GKP approximation is too bad (small $s_\text{GKP}$) or too much loss accumulates (large $L$).
Since increasing the squeezing poses a core experimental challenge,
the smallest value of $s_\text{GKP}$ at which a nonzero SKR can be achieved is of particular interest.
Below $s_\text{GKP} = 10\unitspace\text{dB}$, neither protocol is suitable for generating secret keys.
For both protocols and for every fixed value of $L$, 
we observe that GKP qubits ($D=2$) represent the leading contender for near-term quantum repeaters based on the GKP code.
To some degree, this result is surprising because in the ideal case, 
the SKR per channel use is given by $\log_2(D)$, and increasing the qudit dimension would be beneficial.
In the presence of noise, however, higher-dimensional GKP qudits have the severe disadvantage of decreased error correction capabilities: a $D$-dimensional GKP qudit can only correct displacement errors that are smaller than $\sqrt{{\pi}/{2D}}$ in magnitude.
Only in the regime of very small errors, i.e., where the qubit GKP protocol has almost reached its maximum performance of $\log_2(D) - H(\mathcal{P}) = \log_2(2)-0 = 1.0$,
it is beneficial to employ qutrits ($D=3$) instead of qubits.
To see such benefits at all, we need at least $s_\text{GKP} \gtrsim 18\unitspace\text{dB}$.
For repeater lines of modest lengths of a few ten kilometers, however, larger squeezing levels of $20\unitspace\text{dB}$-$25\unitspace\text{dB}$ are required to compensate for additional loss.
At some value of $L$, loss errors become so severe that only an unrealistically disproportional improvement of $s_\text{GKP}$ could compensate them.
For the one-way protocol in Fig.~\ref{fig:figa}~(a),
qutrits cease to be the optimal option for repeaters longer than a few hundred kilometers, whereas 
the two-way protocol in Fig.~\ref{fig:figa}~(b) 
can still benefit from qutrits even for repeaters exceeding $L= 10,000\unitspace\text{km}$.
For the latter, however, a squeezing level above $30\unitspace\text{dB}$ is required, which will only be available in the long term (if at all).
The reason for the better performance of the two-way protocol is the lower required amplification factor $\sqrt{\eta}^{-1}$ in the usage of the classical post-amplification, as discussed in Sec.~\ref{sec:loss_and_amplification}. 

Finally note that, in our error analysis, we distinguish the cases of even and odd qudit dimensions.
Only if $D$ is even, we can leverage a beneficial linear-optics protocol for the generation of GKP Bell pairs, see App.~\ref{app:error_analysis_bare}.
For very short repeater chains, we indeed observe that GKP qudits with $D$ even outperform those with $D$ odd.
For larger values of $L$, however, loss errors begin to dominate and parameter regions emerge where the optimal SKR is obtained by odd-dimensional GKP qudits.

\subsection{Repeater performance with both GKP and higher-level error correction}

In comparison to the experimental challenge of creating high-quality GKP qudits in the first place,
concatenating multiple of them into a single logical qudit by means of a higher-level QECC is relatively straightforward.
In the following, we study the performance of third-generation quantum repeaters that make use of $\llbracket D, 1, \tfrac{D+1}{2} \rrbracket_D$ quantum polynomial codes ($D\ge 3$ prime), as reviewed in a related context in App.~A of Ref.~\cite{MHKB18}.
The Pauli weight of the stabilizer generators is immense for quantum polynomial codes, which renders them unsuitable for applications in quantum computing.
For quantum repeaters, on the other hand, this is not an issue, as non-destructive measurements of stabilizer operators are not required.
Instead, destructively measuring all qudits individually is sufficient here.
This facilitates syndrome extraction and decoding in a purely classical manner.
Since the distance of a quantum polynomial code is given by $d=\tfrac{D+1}{2}$, 
any collection of errors that affect no more than $\lfloor\tfrac{d-1}{2}\rfloor =  \lfloor \tfrac{D-1}{4} \rfloor$ qudits can be corrected.
For error patterns that affect more qudits than this, 
we assume (as a worst-case approximation) that a uniformly random logical error occurs. 
This maximizes the Shannon entropy $H(\mathcal{P})$ 
and lower bounds the SKR, $\log_2(D)-H(\mathcal{P})$, that would be achieved if a more sophisticated decoder for correcting specific high-weight errors was used.
Thus, it makes sense for us to limit the discussion to prime qudit dimensions where $D-1$ is a multiple of four. 
We defer our derivation of $H(\mathcal{P})$ for this suboptimal decoder to App.~\ref{app:error_analysis_encoded}.
\begin{figure}[t]
	\subfloat[]{\label{fig:skr_GKP_with_L0=100m_squeezing=20dB_coupling=0.99twowaycc}\includegraphics[height=0.3\textwidth]{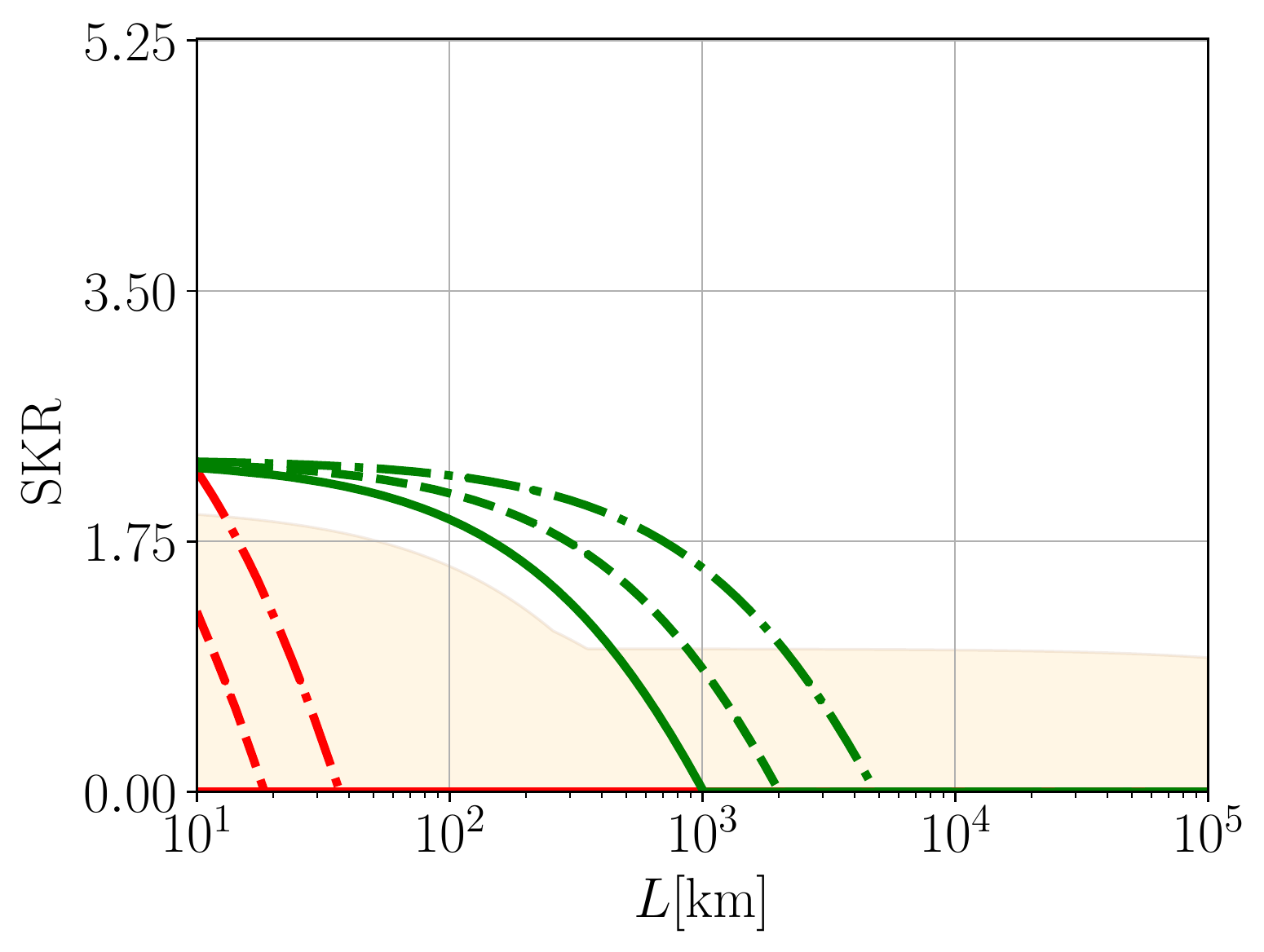}}
	\subfloat[]{\label{fig:skr_GKP_with_L0=100m_squeezing=30dB_coupling=0.99twowaycc}\includegraphics[height=0.3\textwidth]{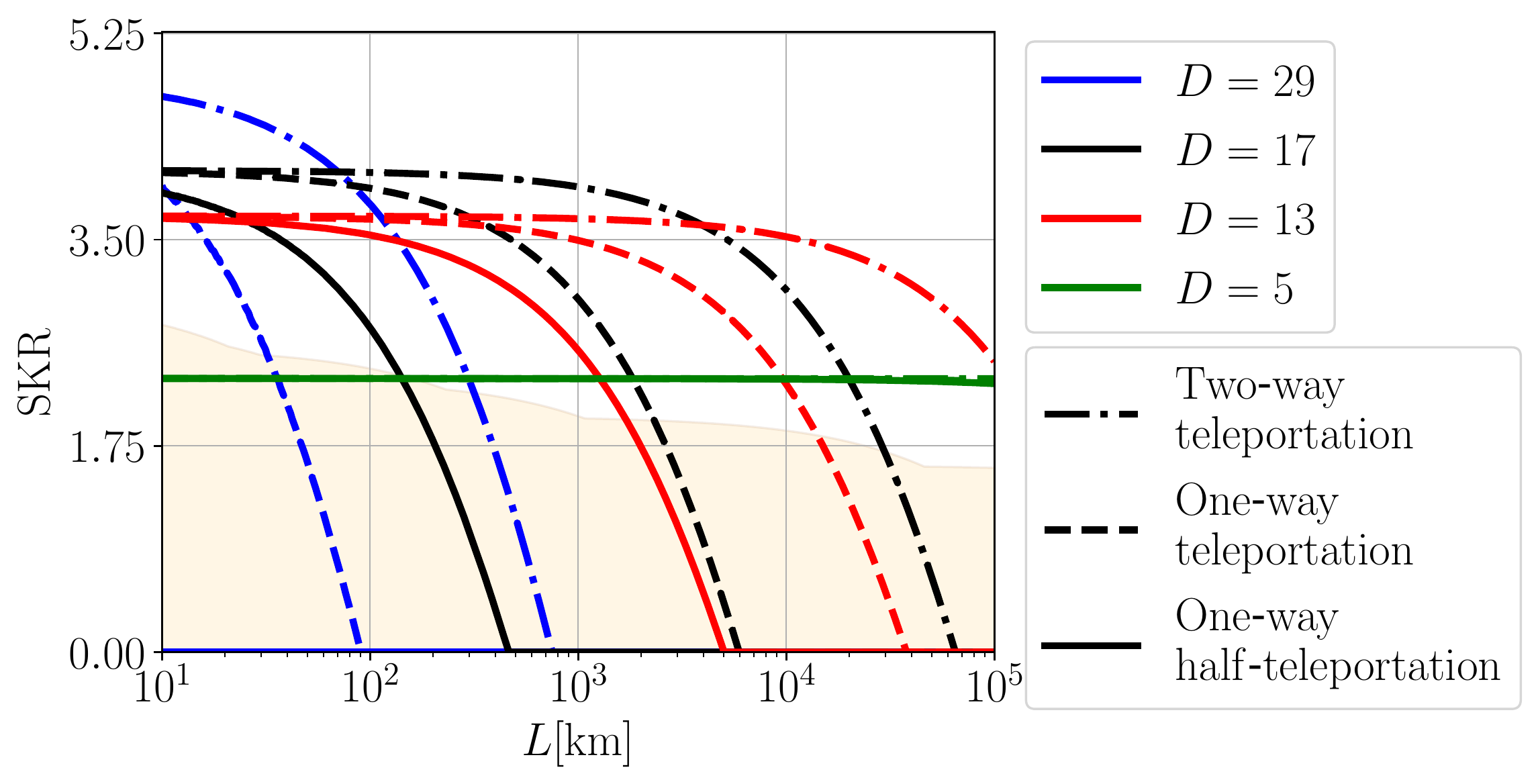}}\\
	\caption{Lower bound on the SKR per logical channel use, $\log_2(D)-H(\mathcal{P})$,  as a function of the total length~$L$ for a quantum repeater line with coupling efficiencies $\eta_\text{c}=99\%$, an intermediate repeater spacing of $L_0 = 100\unitspace\text{m}$, and squeezing levels of (a) $s_\text{GKP} = 20\unitspace\text{dB}$ or (b) $s_\text{GKP}=30\unitspace\text{dB}$.
    The highlighted area shows the achievable SKR per physical channel use of a bare GKP repeater as in Fig.~\ref{fig:figa} (b).
	}
	\label{fig:figb}
\end{figure}

In Fig.~\ref{fig:figb}, we plot the (lower bound on the) SKR per logical channel use as a function of $L$, where $L_0=100\unitspace\text{m}$ is fixed.
For each of the three repeater protocols introduced in Sec.~\ref{sec:protocol}, we show the SKR for $D=5$ (green), $D=13$ (red), $D=17$ (black), and $D=29$ (blue).
For any fixed value of $D$, we again (as in Fig.~\ref{fig:figa}) observe that the two-way teleportation protocol (dash-dotted curve) from Sec.~\ref{sec:2way_postamp} performs best. 
It is followed by the one-way teleportation protocol (dashed curve) from Sec.~\ref{sec:1way_preamp}.
The least-efficient protocol is the one-way half-teleportation protocol (solid curve) from Sec.~\ref{sec:1way_preamp_half}.
We attribute the poor performance of the latter protocol to the fact that it employs only half as many (compared to the other protocols) logical measurements, which facilitate the correction of errors.

Recall from Sec.~\ref{sec:result_bare_gkp} that for bare GKP repeaters, 
the decreased error-correcting capabilities render higher-dimensional qudits unfeasible for near-term applications.
Since the code distance $d=\frac{D+1}{2}$ grows with $D$, 
one could expect that concatenating bare GKP qudits with quantum polynomial codes would turn the tide.
We see that this is not the case: 
for an optimistic but conceivable value of $s_\text{GKP}=20\unitspace\text{dB}$, we see in Fig.~\ref{fig:figb} (a) that only the smallest code with $D=5$ achieves a nonzero SKR for repeater lengths $L>70\unitspace\text{km}$.
To assess the performance of larger codes, we need to assume exorbitant squeezing levels, e.g., $s_\text{GKP}=30\unitspace\text{dB}$ as in Fig.~\ref{fig:figb} (b).
In this scenario, the $\llbracket5,1,3\rrbracket_5$-code operates near its maximum performance of $\log_2(5) \approx 2.3$ for all considered values of $L$.
Depending on the distance $L$, the largest value of $\log_2(D)-H(\mathcal{P})$ is obtained by a different code:
until $L\approx100\unitspace\text{km}$, the $\llbracket 29,1,15 \rrbracket_{29}$-code achieves a value beyond the optimal performance of $\log_2(17)\approx 4.1$ of the $\llbracket 17,1,9 \rrbracket_{17}$-code.
The latter starts to lose performance after a few thousand kilometers, where it falls behind the $\llbracket 13,1,7\rrbracket_{13}$-code.
For comparison, we also show in Fig.~\ref{fig:figb} the performance of the two-way repeater protocol with bare GKP qudits (shaded region), 
where we select the value of $D$ that optimizes the SKR,
as in Fig.~\ref{fig:figa} (b).
For $s_\text{GKP}=20\unitspace\text{dB}$ in Fig.~\ref{fig:figb} (a), bare GKP ququarts ($D=4$) are optimal until  $L \approx 200\unitspace\text{km}$. 
For longer repeaters, too much loss accumulates, and lower-dimensional GKP codes with higher error-correcting capabilities become beneficial:
in a small range of $L$, bare qutrits are the optimal choice, but already for $L\gtrsim300\unitspace\text{km}$, qubits perform best.
As before, this advantage of even dimensions over odd ones is due to improved Bell state availability~\cite{gkp_syndrome}.
For $s_\text{GKP}=30\unitspace\text{dB}$ in Fig.~\ref{fig:figb} (b), 
losses are less of an issue and eight-dimensional GKP qudits are optimal until $L\approx20\unitspace\text{km}$.
For $30\unitspace\text{km} \lesssim L \lesssim 200 \text{km}$, $D=6$ is optimal.
For $1000\unitspace\text{km} \lesssim L \lesssim 50,000 \text{km}$, a bare GKP repeater should operate with $D=4$.

It is important to stress that, from a practical perspective and for the considered parameters, it is not useful to employ higher-level QECCs if the application is quantum key distribution (QKD). 
For example, if $s_\text{GKP}=30\unitspace\text{dB}$ and $L=1000\unitspace\text{km}$, the two-way teleportation protocol with a logical $\llbracket 17,1,9\rrbracket_{17}$-code achieves the largest rate of about four secret bits per logical channel use.
To accomplish this, however, seventeen GKP qudits (entangled in a QECC), i.e., seventeen GKP-encoded and entangled optical modes, need to be transmitted.
With an even lower experimental effort, one could simply transmit in parallel seventeen bare GKP ququarts, i.e., seventeen GKP-encoded but unentangled optical modes, each of which establishes almost two secret bits.
In other words, here the best bare protocol is more efficient than the best higher-level encoded one by a factor of about $8.5$.

\subsubsection{Optimal choice of the repeater spacing}

\begin{figure}[t] \centering
	\subfloat[]{\label{fig:skr_VS_L_with_L=2000km_20_squeezing0.999_coupling_erasure_twoway}\includegraphics[height=0.275\textwidth]{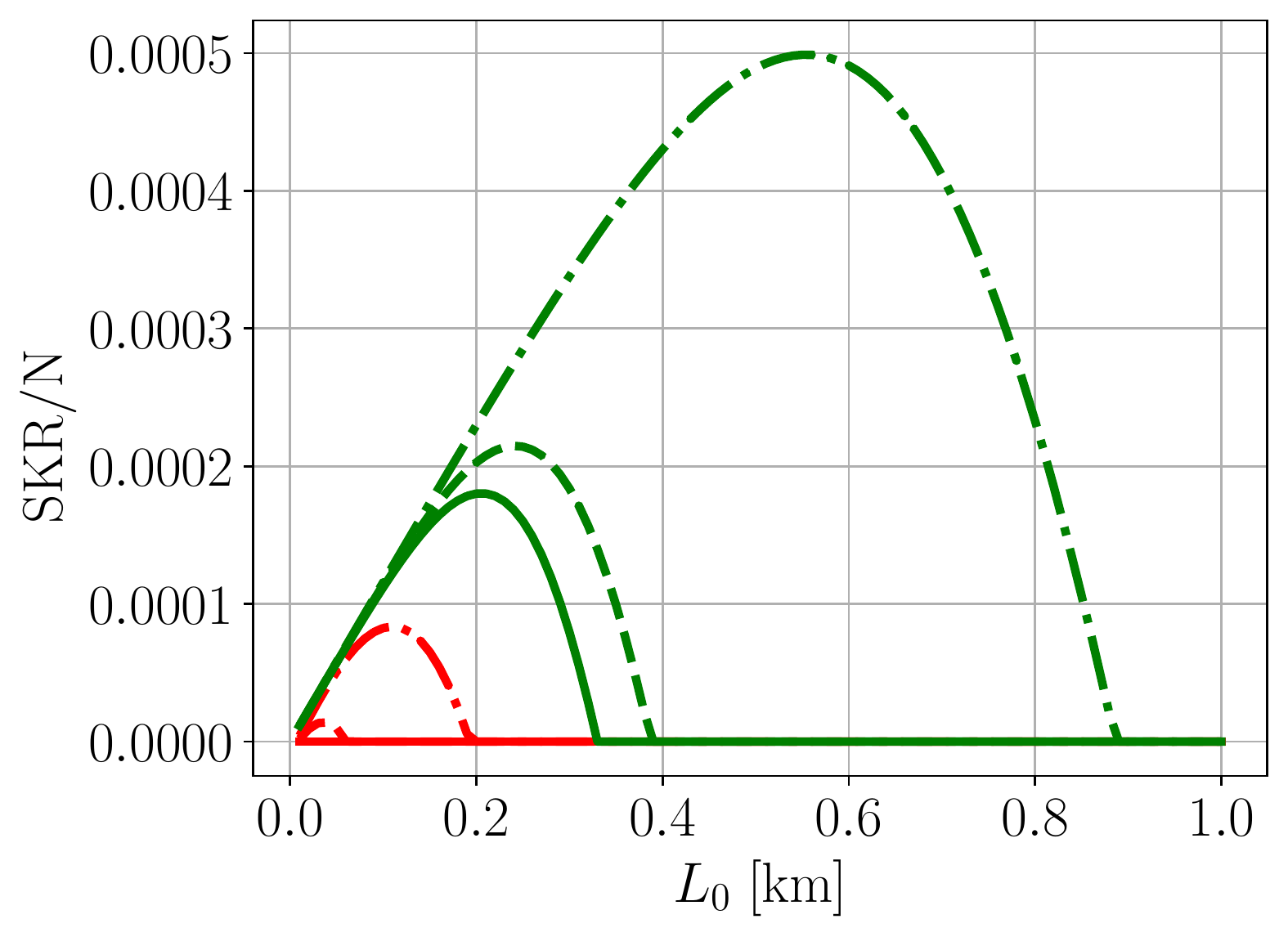}}
    \hspace{5mm}
	\subfloat[]{\label{fig:skr_VS_L_with_L=2000km_30_squeezing0.999_coupling_noerasure_twoway}\includegraphics[height=0.275\textwidth]{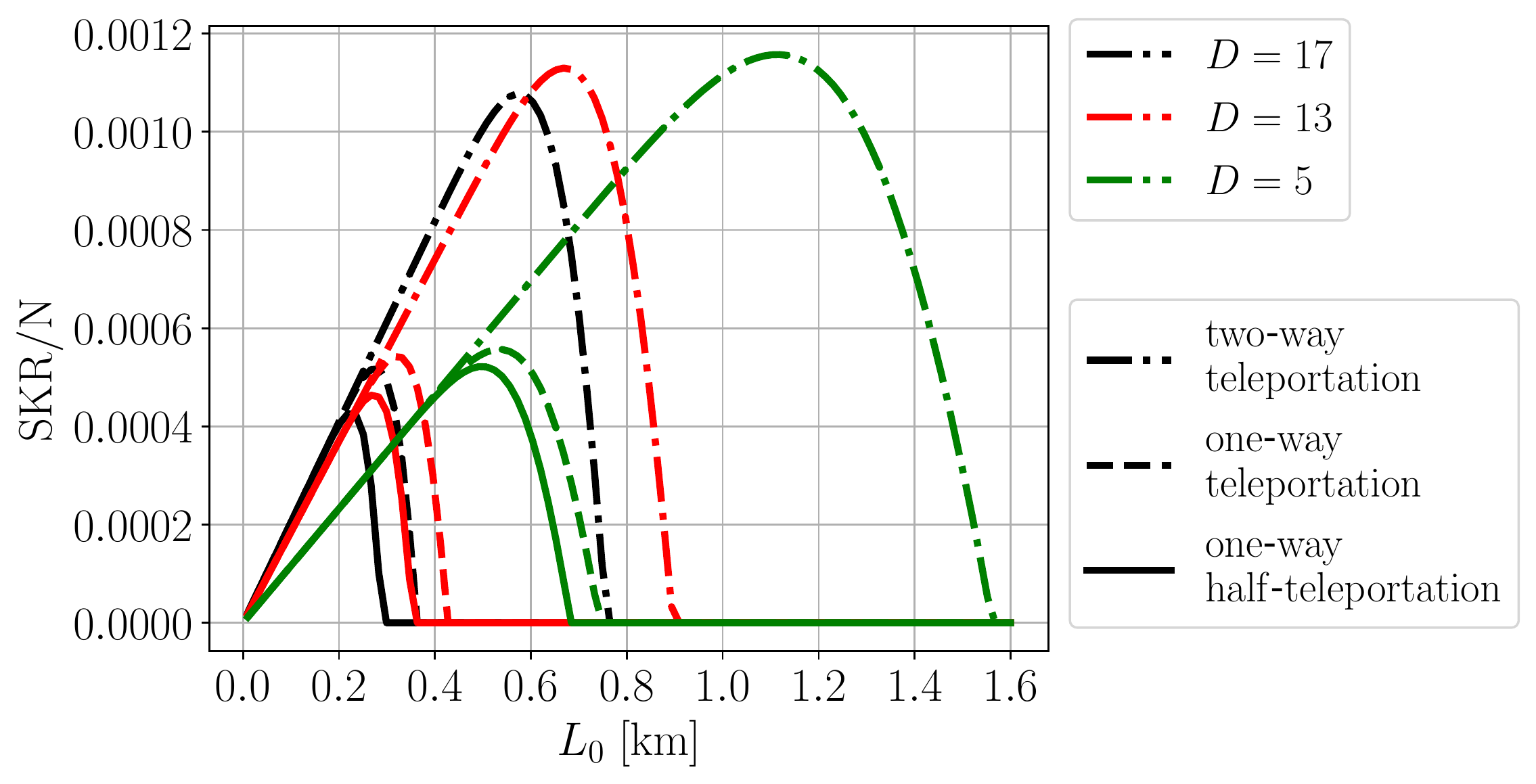}}\\
	\caption{Lower bound on the SKR per logical channel use, $\log_2(D)-H(\mathcal{P})$, normalized by the number $N$ of repeater segments ($N-1$ repeater stations) as a function of the repeater spacing $L_0$ for a repeater line of total length $L = N  L_0 = 2000\unitspace\text{km}$, coupling efficiencies $\eta_\text{c}=99.9\%$, and squeezing levels of (a) $s_\text{GKP} = 20\unitspace\text{dB}$ or (b) $s_\text{GKP}=30\unitspace\text{dB}$.
	}
	\label{fig:figc}
\end{figure}

In our discussion of Fig.~\ref{fig:figb}, we have pointed out that no practical benefit is to be expected when switching from bare GKP qudits to a higher-level QECC 
if the repeater spacing is fixed to $L_0 = 100\unitspace\text{m}$.
This raises the question of how the choice of $L_0$ influences this conclusion.
Since implementation cost scales with the total number $N = L/L_0$ of repeater stations, here we focus on SKR$/N$ as a figure of merit.
In a commercial setting, SKR$/N$ is roughly proportional to the return on investment.
In Fig.~\ref{fig:figc}, we plot SKR$/N$ as a function of $L_0$ for a quantum repeater line of fixed length $L= 2000\unitspace\text{km}$.
The colors and line styles have the same meaning as in Fig.~\ref{fig:figb}.
This time, we assume a more optimistic value of $\eta_\text{c} = 99.9\%$, which benefits higher-level QECCs.
Despite this optimistic assumption, we still find that (for QKD) bare GKP qudits outperform those encoded into quantum polynomial codes.
For example, for $s_\text{GKP}=20\unitspace\text{dB}$ in Fig.~\ref{fig:figc} (a), the $\llbracket5,1,3\rrbracket_5$-code performs best among the quantum polynomial codes and reaches the optimal value of SKR/$N$ at a repeater spacing of $L_0 \approx 0.55\unitspace\text{km}$.
For this optimal repeater configuration, the $\llbracket5,1,3\rrbracket_5$-code can generate approximately 1.8 secret bits by transmitting five GKP ququints ($D=5$).
In the same setting, one can generate almost 5.0 secret bits by transmitting five bare GKP qubits (not shown).
The same behavior is observed for $s_\text{GKP} = 30\unitspace \text{dB}$ in Fig.~\ref{fig:figc} (b), 
where the $\llbracket5,1,3\rrbracket_5$-code now 
achieves approximately 2.0 secret bits per logical channel use at the optimal operating point of $L_0 \approx 1.1\unitspace\text{km}$.
In the same setting, transmitting five bare GKP qutrits would generate more than 5.6 secret bits.

From Fig.~\ref{fig:figc}, we can also infer the maximal repeater spacing at which the secret-key rate drops to zero.
For the considered parameters, the best higher-level encoded protocol, i.e., the two-way protocol from Sec.~\ref{sec:2way_postamp} with the $\llbracket5,1,3\rrbracket_5$-code and $s_\text{GKP}=30\unitspace\text{dB}$, 
is operational for all values of $L_0<1.5\unitspace\text{km}$, however, $L_0 \approx 1.1\unitspace\text{km}$ is most effective.
For the one-way protocols from Secs.~\ref{sec:1way_preamp} and~\ref{sec:1way_preamp_half}, the $\llbracket5,1,3\rrbracket_5$-code already fails for $L_0 \approx 0.7\unitspace\text{km}$.
As expected, we find that better repeaters (larger $s_\text{GKP}$, smaller $D$) allow for a larger repeater spacing.

\subsubsection{Identifying and overcoming noise bottlenecks}

\begin{figure} \centering
    \subfloat[]{\label{fig:squeezing_coupling_phasespace_spacing=500m}
	\begin{minipage}{0.45\textwidth}
	\vspace*{-41.9mm}
	
	\includegraphics[width=\textwidth]{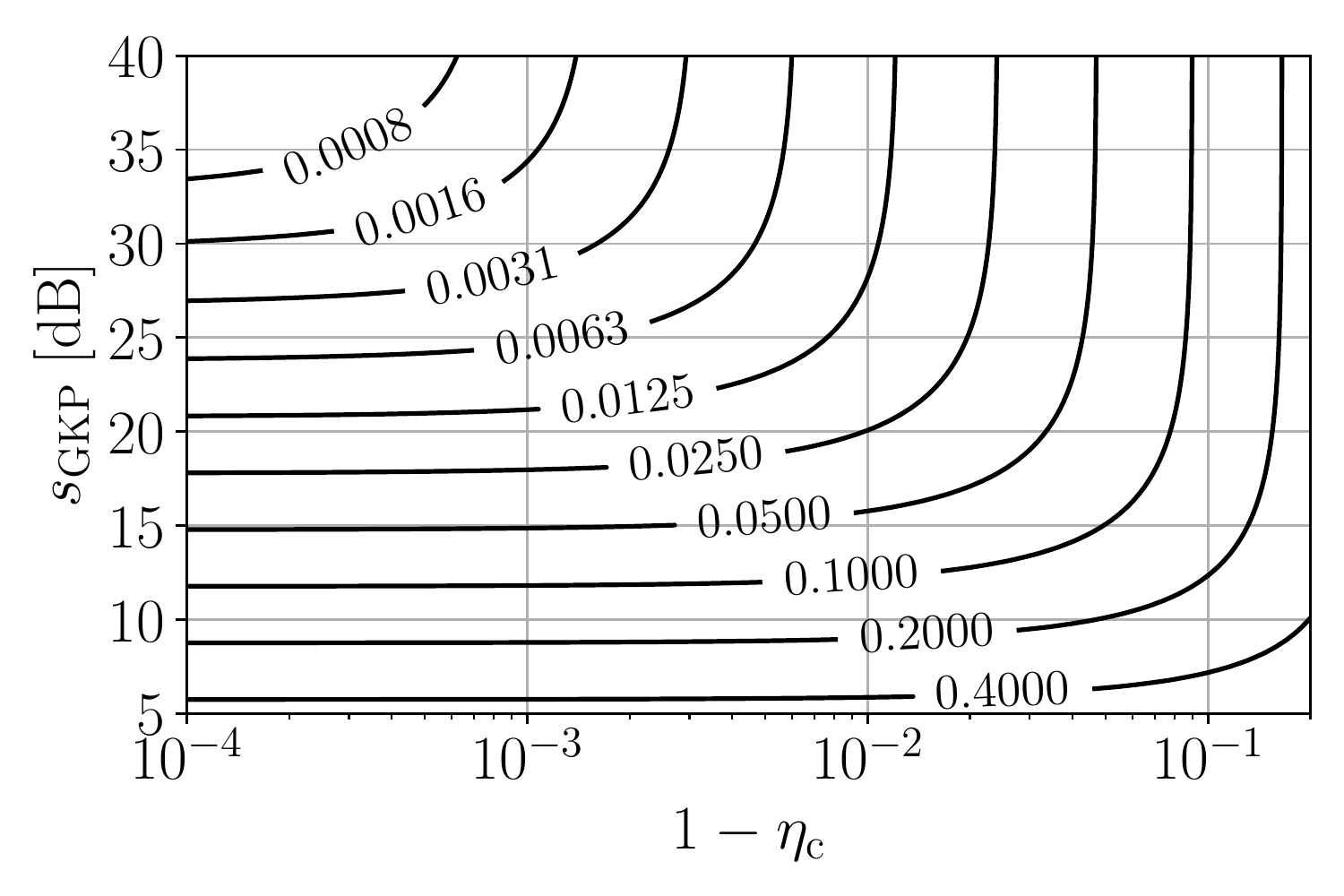}
	\end{minipage}
	}
    \hspace{10mm}
	\subfloat[ ]{\label{fig:skr_GKP_with_internal_noise_L=5000km_spacing0.5km_twowaycc}\includegraphics[width=0.45\textwidth]{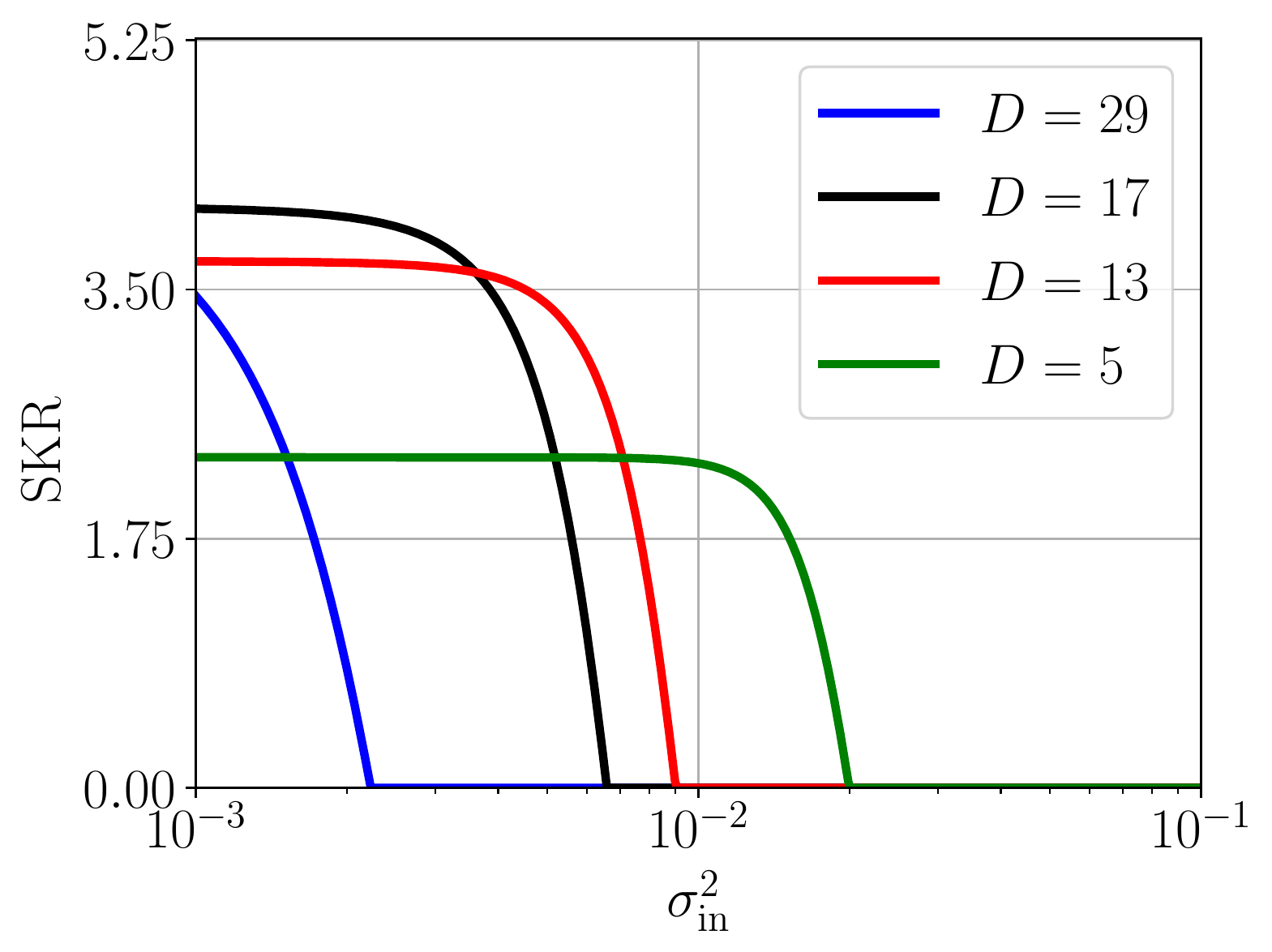}}
	
	\caption{(a) Variance $\sigma_\text{in}^2$ of Gaussian noise effectively affecting a physical GKP qudit after it has been coupled into fiber as a function of the squeezing level $s_\text{GKP}$ and coupling efficiency $\eta_\text{c}$. 
    (b) Lower bound on the SKR per logical channel use, $\log_2(D)-H(\mathcal{P})$, as a function of $\sigma_\text{in}^2$ for a quantum repeater line with a total transmission distance of $L = 5000\unitspace\text{km}$ and a repeater spacing of $L_0 = 500\unitspace\text{m}$, where the two-way protocol in combination with a $\llbracket D, 1, \tfrac{D+1}{2}\rrbracket_D$ quantum polynomial code is utilized.}
	\label{fig:figd}
\end{figure}

Before one takes a great effort of building a quantum repeater based on GKP qudits,
it is important to ensure that the experimental building blocks work sufficiently well.
There are multiple components for which improvements might be beneficial or even necessary. 
Thus, it is important to identify and remove the noise bottleneck, which would otherwise diminish the performance.
We distinguish three error mechanisms: input noise, fiber channel losses, and imperfect homodyne measurements. 
Since measurements work comparatively well and we have already discussed the impact of fiber loss,
here we focus on input noise that arises from approximate GKP state preparation and  coupling losses.
As explained in Sec.~\ref{sec:noise},
both processes can be modeled by Gaussian noise.
Errors propagate through the circuit and eventually accumulate on individual measurement results in the repeater stations, which for the two-way post-amplification protocol from Sec.~\ref{sec:2way_postamp} can be described by a Gaussian channel with variance 
\begin{align} \label{eq:variance_input}
    \sigma_\text{in}^2 = 3 \sigma_\text{sq}^2 + \frac{1-\eta_\text{c}}{\eta_\text{c}} \exp\left(\frac{L_0}{2L_\text{att}}\right).
\end{align}
Indeed,
there are three sources from which GKP state preparation errors can propagate to the measurements, which leads to the first term in Eq.~\eqref{eq:variance_input}.
The second term in Eq.~\eqref{eq:variance_input} accounts for coupling losses:
since the variance (incorporating both coupling and fiber channel losses) of a length-$L_0$ link 
in the two-way protocol is given by 
$(\eta_\text{c}\exp(-{L_0}/{2L_\text{att}}))^{-1}-1$,
the noise difference between a link with coupling errors and without is given by
\begin{align}
((\eta_\text{c}\exp(-\tfrac{L_0}{2L_\text{att}}))^{-1}-1 ) 
-
((\exp(-\tfrac{L_0}{2L_\text{att}}))^{-1}-1) 
=
     \frac{1-\eta_\text{c}}{\eta_\text{c}} \exp\left(\frac{L_0}{2L_\text{att}}\right).
\end{align}

In Fig.~\ref{fig:figd} (a), we plot $\sigma_\text{in}^2$ as a function of both $s_\text{GKP}$ and $\eta_\text{c}$.
Recall that $\sigma_\text{sq}^2$ and $s_\text{GKP}$ can be converted into each other via Eq.~\eqref{eq:squeezing_parameter}.
Here, we assume a repeater spacing of $L_0 = 500\unitspace\text{m}$, however, the situation remains almost unchanged if $L_0$ takes any other value between $1\unitspace\text{m}$ and $1\unitspace\text{km}$.
The contour lines in Fig.~\ref{fig:figd} can be used to infer whether one should work on improving $s_\text{GKP}$ or $\eta_\text{c}$: 
since moving along a contour line does not improve performance, 
a series of improvements should instead correspond to a path orthogonal to the contour lines.
For example, for $s_\text{GKP}=6\unitspace\text{dB}$ and $\eta_\text{c} = 0.99$, we have $\sigma_\text{in}^2 \approx 0.4$, which can be reduced to $\sigma_\text{in}^2 \approx 0.2$ if the GKP approximation is improved to  $s_\text{GKP}=9\unitspace\text{dB}$; increasing $\eta_\text{c}$, on the other hand, would not help at all.
Conversely, if coupling losses dominate, e.g.,  $s_\text{GKP}=30\unitspace\text{dB}$ and $\eta_\text{c} = 0.92$, the variance $\sigma_\text{in}^2 \approx 0.1$ can be reduced by a factor of two if coupling efficiencies are improved to $\eta_\text{c} = 0.97$; increasing $s_\text{GKP}$, however, would show no significant effect here.

In Fig.~\ref{fig:figd} (b), we depict the influence of $\sigma_\text{in}^2$ on the SKR obtained with the two-way protocol from Sec.~\ref{sec:2way_postamp} 
for an error-corrected quantum repeater line with $L=5000\unitspace\text{km}$, $L_0 = 500 \unitspace\text{m}$, and a $\llbracket D, 1, \tfrac{D+1}{2}\rrbracket_D$-code.
Here, each physical GKP qudit in every repeater station is affected by a Gaussian channel with variance $\sigma_\text{in}^2$.
Note that also the effect of imperfect homodyne measurements can be inferred from Fig.~\ref{fig:figd} (b) if a corresponding variance term $\sigma_\text{meas}^2$ is added to $\sigma^2_\text{in}$.
As before, we find that a larger value of $D$ both allows for a larger SKR per logical channel use in the low-noise regime and for a smaller noise level to be tolerated before the SKR drops to zero.
We also observe that the parameter range of $\sigma_\text{in}^2$ where the SKR drops from its optimal value to zero is alarmingly small.
This effect is most pronounced for the $\llbracket 5,1,3\rrbracket_5$-code, which has almost optimal performance until $\sigma_\text{in}^2\approx 0.01$ but already for $\sigma_\text{in}^2\approx 0.02$ its SKR is equal to zero.
This showcases that moderate improvements can have a huge impact if they address a noise bottleneck.

\subsubsection{Leveraging lower-level syndrome information to improve higher-level error correction}

So far, we have independently treated the error correction procedures of lower-level GKP qudits and the higher-level QECC.
More precisely, we assumed that, in the first step, displacement errors on the physical GKP qudits are removed.
This may or may not result in a logical GKP qudit error.
Then, in a second step, the higher-level $\llbracket n, 1,d \rrbracket_D$-code deals with potential errors on the GKP qudits:
the correction succeeds if the number of errors with unknown locations is not larger than $\frac{d-1}{2}$.
In this final subsection, we investigate the more general case, where the location of some of the errors are known.
The modified error correction routine succeeds whenever $t_\text{k} +2t_\text{u} <d$, where $t_\text{k}$ and $t_\text{u}$ denote the number of errors with known and unknown locations, respectively.
To obtain some information about error location, one can exploit the continous, ``analog’’ results of the homodyne measurements in the repeater stations~\cite{gkp_analoginfo, toric-gkp}.
If a displacement error of the form $\exp(\text{i}\epsilon\hat{p})$ occurs, the homodyne measurement of $\hat{q}$ reveals the value of $\epsilon$ modulo $\sqrt{2\pi/D}$, which we call analog GKP syndrome.
In particular, every displacement error with  $|\epsilon| < \sqrt{{\pi}/{2D}}$ can be corrected.
The probability of successful error correction is large if $\epsilon$ is small.
When an error of magnitude $\epsilon \approx \sqrt{{\pi}/2D}$ occurs, 
however, the situation is less clear.
Borrowing ideas from Ref.~\cite{fukui2021efficient}, we introduce a discarding parameter $\gamma \in [0,1]$,
and treat any instances of $\epsilon$ which are closer than $\sqrt{\pi/2D}(1-\gamma)$ from the boundary of two bins
as an erasure error with a known location.
In the case $\gamma=1$, we do not discard any qudits, which corresponds to the strategy considered so far.
The other extreme, $\gamma = 0$, corresponds to the absurd approach where all qudits are always discarded.

The advantage of this modification is that, 
for every qudit that is not discarded, 
the probabilities for errors (with unknown locations) are improved from Eq.~\eqref{eq:pauli_shift_probability_gauss} to
\begin{align} 
 P^{(\gamma)}_\text{sq}(X^k,\sigma^2) \propto 
 \sum_{j\in\mathbb{Z}}\int
     _{\sqrt{\frac{2\pi}{D}}(jD+k-\frac{\gamma}{2})}
     ^{\sqrt{\frac{2\pi}{D}}(jD+k+\frac{\gamma}{2})}
     \frac{1}{\sqrt{2\pi \sigma^2}}\exp\left(-\frac{q^2}{2\sigma^2}\right)dq,
\end{align}
where the proportionality constant follows from  $\sum_{k=0}^{D-1}  P^{(\gamma)}_\text{sq}(X^k,\sigma^2) = 1$.
This improvement comes at the expense that we have to introduce an erasure error with probability
\begin{align}
   p_\text{discard} = 1- \sum_{k=0}^{D-1} 
 \sum_{j\in\mathbb{Z}}\int
     _{\sqrt{\frac{2\pi}{D}}(jD+k-\frac{\gamma}{2})}
     ^{\sqrt{\frac{2\pi}{D}}(jD+k+\frac{\gamma}{2})}
     \frac{1}{\sqrt{2\pi \sigma^2}}\exp\left(-\frac{q^2}{2\sigma^2}\right)dq,
\end{align}
however, we can still exploit our knowledge about the location of this error.

Denote the probability that a single GKP qudit is free of errors by $p_0 = P^{(\gamma)}_\text{sq}(X^0, \sigma^2)$.
Then,  the condition $t_\text{k} +2t_\text{u} <d$ and basic combinatorics leads to the probability of a failed error correction attempt
\begin{align} \label{eq:p_fail}
    p_\text{fail}(\gamma) = 1-\sum_{t_\text{k}=0}^{d-1}\binom{n}{t_\text{k}} p_\text{discard} ^{t_\text{k}} (1-p_\text{discard} )^{n-t_\text{k}} 
    \sum_{t_\text{u}=0}^{t_\text{u,max}} \binom{n-t_\text{k}}{t_\text{u}} p_0^{n-t_\text{k}-t_\text{u}} (1-p_0)^{t_\text{u}},
\end{align}
where $t_\text{u,max} = \left\lfloor ({d-t_\text{k}-1})/{2} \right\rfloor$ is the maximal number of correctable errors with unknown locations, assuming that $t_\text{k}$ erasures occurred, and $n$ is the number of physical GKP qudits.

\begin{figure}
	\centering
	\includegraphics[width=0.6\linewidth]{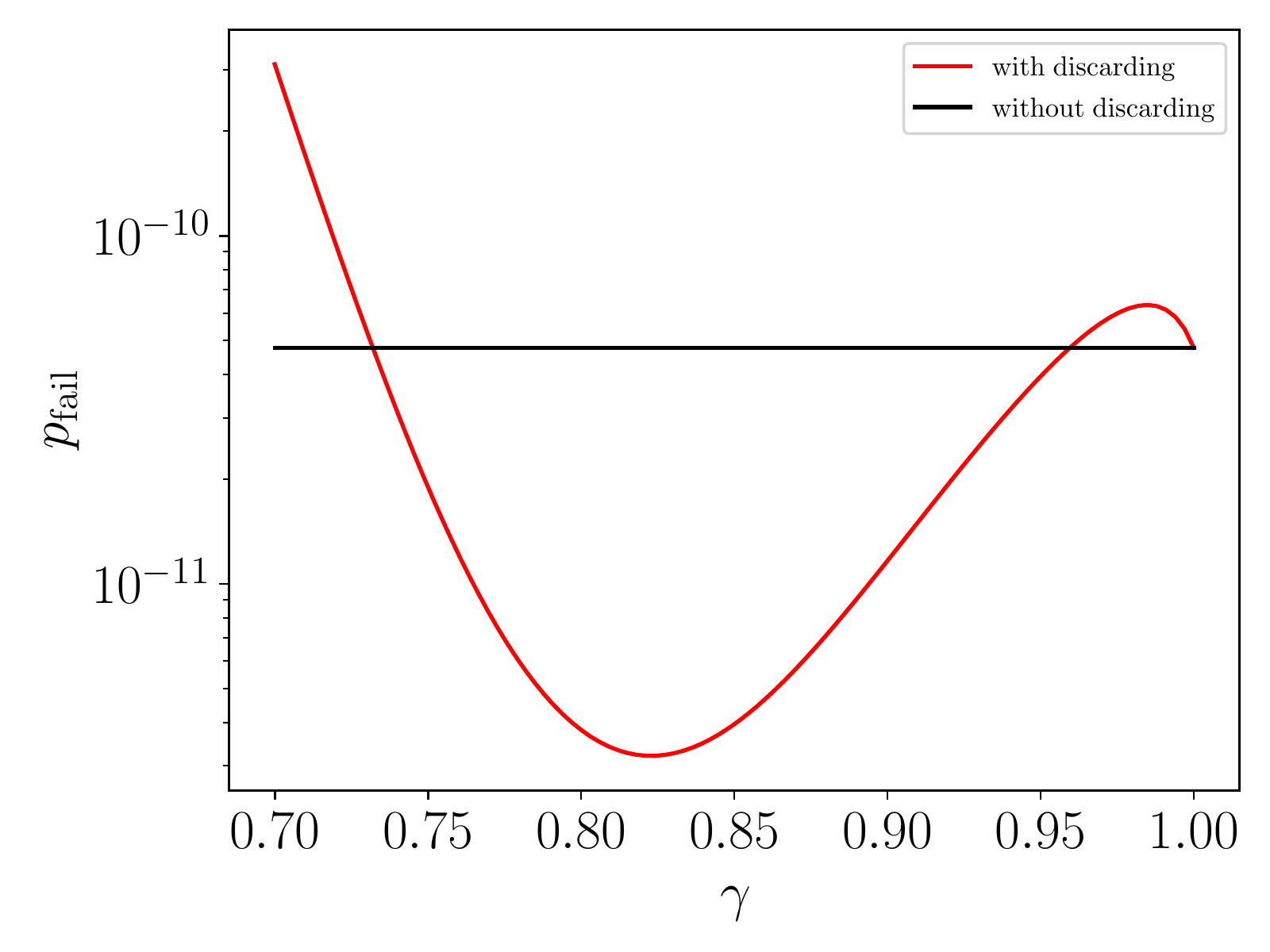}
	\caption{Failure probability $p_\text{fail}$ for decoding the result of a logical measurement for a 
    $\llbracket 13, 1, 7\rrbracket_{13}$-code as a function of the discarding parameter $\gamma$.
    Each physical GKP qudit is subject to Gaussian noise with variance $\sigma^2=0.01$. }
	\label{fig:erasureplotn13}
\end{figure}

In Fig.~\ref{fig:erasureplotn13}, we show how the logical failure rate (red) depends on the discarding parameter $\gamma$ for a $\llbracket13,1,7\rrbracket$-code.
For each physical GKP qudit, we assume that all noise combined (stemming, e.g., from GKP approximation, coupling, or transmission) corresponds to a fairly small but finite variance $\sigma^2=0.01$ of the overall Gaussian noise channel.
For $\gamma = 1$, i.e., without discarding (black), the failure rate has a remarkably low value of $p_\text{fail} \approx 5\times 10^{-11}$, which is due to the low level of noise and the high error-correcting distance of $d=7$.
We observe a local minimum at $\gamma_\text{opt} \approx 0.82$,  where 
the failure rate is improved by more than an order of magnitude to $p_\text{fail}(\gamma_\text{opt}) \approx 3\times 10^{-12}$.
If $\gamma$ is decreased below $\gamma_\text{opt}$, we begin to introduce more erasures than the QECC can deal with, and the failure rate increases.
On the other hand, if $\gamma$ is increased above $\gamma_\text{opt}$, 
then the error rates $P^{(\gamma)}_\text{sq}(X^k,\sigma^2)$ start to deteriorate.
This causes an increasing amount of errors with unknown locations and leads to the rise in $p_\text{fail}$.
A curious effect in Fig.~\ref{fig:erasureplotn13} is that the performance 
in the seldom-discarding regime ($\gamma > 0.96$) is worse than in the never-discarding case ($\gamma = 1$).
We attribute this to the fact, that for $0.96 < \gamma < 1.0$, those cases dominate where only a single erasure error is introduced, i.e., $t_\text{k}=1$ and the number of correctable errors with unknown locations is decreased to $t_\text{u,max} = 2$.
At the same time, the error probabilities $P^{(\gamma)}_\text{sq}(X^k,\sigma^2)$ are only slightly improved because they continuously depend on $\gamma$.
Thus, the performance is worse than for the naive approach with $\gamma=1$, i.e., $t_\text{k}=0$ and $t_\text{u,max}=3$.

\section{Conclusion and outlook}
\label{sec:4}
In this paper, we have analyzed the performance of third-generation quantum repeaters that operate with higher-dimensional GKP qudits.
We have focused on the GKP square lattice and also considered concatenations with quantum polynomial codes.

The missing component that is currently holding back an experimental realization of such repeaters is efficient sources of high-quality GKP states.
Once such sources are available, however, 
there will be no need for quantum memories or classical two-way communication.
Therefore, the achievable repetition rates will only be limited by fast optical elements for the local processing of GKP qudits.

Our initial motivation for the present investigation was that,
at a first glance,
GKP qudits and quantum polynomial codes seem like a perfect match in the context of quantum repeaters:
GKP states can be encoded into photons, which is crucial for repeaters;
polynomial codes achieve the singleton bound at the expense of high-weight stabilizers, which is a problem for quantum computation but not for quantum repeaters;
polynomial codes require higher-dimensional qudits, which the GKP encoding has to offer.
However, our study revealed that the decreased lower-level error-correcting capabilities of higher-dimensional GKP qudits severely limit their potential benefits.
While this finding might disappoint to a certain extent, it is somewhat good news for experimentalists.
Indeed, the most promising GKP repeater protocol identified in this work is also the one, which is the easiest to implement.

Our recommendation for a first experimental target is a repeater protocol (Sec.~\ref{sec:2way_postamp}) that makes use of two-dimensional GKP qubits.
Admittedly, these qubits will require challenging squeezing levels beyond $10\unitspace\text{dB}$.
However, the identified protocol has the advantage of readily available syndrome measurements based on balanced beam splitters and homodyne measurements alone. 
Furthermore, this protocol is compatible with rescaling the GKP lattice in classical software, 
whereas other protocols would require optical amplifiers to compensate for the loss.

We also found that, in the medium-to-long term,
when squeezing levels above $20\unitspace\text{dB}$ will be available,
the error-correcting capabilities of bare GKP qutrits will suffice to outperform GKP qubits for meaningful repeater lengths.
Only in the very long term, if squeezing levels around $30\unitspace\text{dB}$ can possibly be reached,
we expect some benefit from concatenating multiple GKP qudits using quantum polynomial codes,
however, only for tasks like entanglement distribution where utmost fidelities are important.
For the application of quantum key distribution, on the other hand,
our analysis showed that it is typically more cost-effective to operate bare GKP qudit repeaters instead. With regards to potential experimental realizations, a useful feature
of the case with bare GKP qudits is that the necessary GKP two-qudit
Bell pair for teleportation-based syndrome detection and error correction
can often be directly created by applying a balanced beam splitter upon two
suitable, individual single-mode GKP/grid states~\cite{WalshePRA2020,gkp_syndrome}. In case that the concatenation with the higher-level code is employed,
for potential, future high-fidelity quantum network applications,
the complete syndrome information of the QECC
can still be obtained in one linear-optics step
with no need for any online squeezing operations and also
with no need for separating the physical GKP qudit from
the higher-level code's syndrome measurements and adding extra
GKP ancilla states for the higher-code detections.
This only works, however, provided a suitable logical, higher-level
Bell pair is available~\cite{gkp_syndrome}.

In this paper, we have focused on the cases of single GKP qudits and multiple GKP qudits that are concatenated by means of a higher-level qudit stabilizer code.
This is, however, not the only possibility that can be envisioned.
An interesting open research direction is to study the performance of multi-mode GKP codes that do \emph{not} arise as a concatenation of physical GKP states and a higher-level stabilizer code~\cite{Conrad2022gottesmankitaev, conrad_good_gkp_2023,PRXQuantum.3.010335,lin2023closest,gkp_syndrome}.
For such an analysis, theoretical insights about multi-mode Gaussian channels might become important~\cite{Caruso_2008}.
Moreover, one could analyze how bosonic encodings other than GKP perform in a quantum repeater setting, e.g., cat codes~\cite{cat_codes,li2022memoryless}, spherical codes~\cite{spherical_codes}, etc.~\cite{bosonic_codes,PhysRevX.10.011058}.

\begin{acknowledgments}	
FS and PvL acknowledge financial support from the BMBF in Germany via the projects QR.X, QuKuK, and PhotonQ and
the BMBF/EU for support via the project QuantERA/ShoQC.
DM acknowledges financial support from the BMBF in Germany via the projects RealistiQ, QR.X, and QSolid.

\end{acknowledgments}  

\newpage 
\appendix
\section{Error analysis of bare GKP repeaters}
\label{app:error_analysis_bare}

We begin our error analysis by reviewing how Gaussian displacement errors of the form $\exp(\epsilon \hat q_i)$ and $\exp(\epsilon \hat p_i)$, where $\epsilon\in \mathbb R$ is the error magnitude,
propagate across $\CSUM$- and $\CPhase$-gates.
The $\CSUM$-gate, $\exp\left(-\text{i} \hat{q}_1\hat{p}_2\right)$
acts as $CX_{1,2} = \sum_{k=0}^{D-1} \ket{k}\bra{k}_1 \otimes X_2^k$ on GKP qudits,
while the $\CPhase$-gate, $\exp\left(\text{i} \hat{q}_1\hat{q}_2\right)$, 
implements $CZ_{1,2}=  \sum_{k=0}^{D-1} \ket{k}\bra{k}_1 \otimes Z_2^k$~\cite{gkp}.
Hereby, $X = \sum_{k=0}^{D-1} \ket{k+1 \text{ mod }D} \bra{k}$ and $Z=\sum_{k=0}^{D-1} (\text{e}^{2\pi \text{i}/D})^k \ket{k}\bra{k}$ denote the unitary generalizations of the qubit Pauli $X$- and $Z$-gates to the case of $D$-dimensional qudits.
It is well known that single-qudit Pauli errors are propagated across $CX$- and $CZ$-gates via
\begin{align}\label{eq:propagation_discrete}
\begin{split}
    CZ_{1,2} X_1 & = X_1  Z_2\, CZ_{1,2}, \\
    CX_{1,2} X_1 & = X_1  X_2\, CX_{1,2},  \\ \text{and } \
    CX_{1,2} Z_2 & = Z_1^\dagger Z_2\, CX_{1,2},
\end{split}
\end{align}
see, e.g., Refs.~\cite{gkp, MHKB18}.
The error propagation rules of Eq.~\eqref{eq:propagation_discrete} have their bosonic analogs: applying the Baker-Campbell-Hausdorff formula yields
\begin{align}\label{eq:propagation}
\begin{split}
\exp\left(\text{i} \hat{q}_1\hat{q}_2\right)\exp\left(\text{i} \hat{p}_1\right)
&=\exp\left(\text{i} (\hat{p}_1-\hat{q}_2)\right)\exp\left(\text{i} \hat{q}_1\hat{q}_2\right), \\
\exp\left(-\text{i} \hat{q}_1\hat{p}_2\right)\exp\left(\text{i} \hat{p}_1\right)
&=\exp\left(\text{i}( \hat{p}_1+\hat{p}_2)\right)\exp\left(-\text{i} \hat{q}_1\hat{p}_2\right), \\  \text{and } \
\exp\left(-\text{i} \hat{q}_1\hat{p}_2\right)\exp\left(\text{i} \hat{q}_2\right)
&=\exp\left(\text{i}( \hat{q}_2-\hat{q}_1)\right)\exp\left(-\text{i} \hat{q}_1\hat{p}_2\right).
\end{split}
\end{align} 

In the two repeater protocols from Sec.~\ref{sec:2way_postamp} and~\ref{sec:1way_preamp},
every repeater station is responsible for performing a Bell measurement.
This is achieved by a beam splitter, followed by two homodyne measurements.
For both of these homodyne measurements, the results are post-processed (binned) into a measurement outcome of the GKP qudit.
Errors on the GKP qudit lead to errors on the measurement outcomes.
The latter can be described by a Pauli error channel $\mathcal{P}_\text{sq}(X,\sigma^2)$, as in Eq.~\eqref{eq:pauli_shift_probability_gauss}, where the variance $\sigma^2$ comprises all Gaussian noise contributions that have propagated to the measurement device.
As discussed in Sec.~\ref{sec:noise}, we take the following  error sources into account:
\begin{itemize}
    \item Loss that arises when GKP qudits are coupled into an optical fiber. 
    The resulting coupling efficiency is denoted by $\eta_\text{c}$.
    \item Loss that arises during transmission. 
    If the traveling distance is $L_0$, the associated transmittance is given by $\eta = \exp(-L_0/L_\text{att})$, where $L_\text{att}$ is the attenuation distance.
    \item Unavoidable approximation errors of square GKP qudits. These are modeled by a Gaussian channel of variance $\sigma_\text{sq}^2$.
\end{itemize}
Since beam splitters and homodyne measurements only require passive linear optical elements,
we assume they work perfectly.
Similarly, we ignore errors stemming from Gaussian elements, i.e., from $\CSUM$- and $\CPhase$-gates.

For the two-way teleportation protocol from Sec.~\ref{sec:2way_postamp},
transmission and coupling losses lead to a Gaussian error channel with variance $\frac{1}{\eta_c \sqrt{\eta}}-1$, see Sec.~\ref{sec:loss_and_amplification}.
Furthermore, there are three GKP state preparations in the causal light cone of any given measurement. 
All in all, this amounts to a final variance of 
$\sigma_\text{2-way}^2 = 3\sigma_\text{sq}^2+\frac{1}{\eta_\text{c} \sqrt{\eta}}-1$.

For the one-way teleportation protocol from Sec.~\ref{sec:1way_preamp}, the only difference is that the Gaussian error channel arising from losses now has a variance of
$1-\eta_\text{c}\eta$, see Sec.~\ref{sec:loss_and_amplification}.
Therefore, the final variance is given by 
$\sigma_\text{1-way}^2 = 3\sigma_\text{sq}^2 + 1-\eta_\text{c}\eta$.

If $D$ is even, it is possible to directly generate a two-qudit GKP Bell pair
by applying a balanced beam splitter to two grid states~\cite{WalshePRA2020, gkp_syndrome}.
Unlike general Gaussian transformations,
this linear optical transformation does not amplify the noise.
In consequence, the above variances are improved to 
$\sigma_\text{2-way}^2 = 2\sigma_\text{sq}^2+\frac{1}{\eta_\text{c} \sqrt{\eta}}-1$ and
$\sigma_\text{1-way}^2 = 2\sigma_\text{sq}^2 + 1-\eta_\text{c}\eta$.

On the physical level, every Bell measurement is comprised of two homodyne measurements.
Errors on the measurement of one quadrature effectively propagate into $X$-errors on Bob's qudits, while those of the other quadrature lead to $Z$-errors.
By symmetry, the final probability distributions for $X$- and $Z$-errors coincide, and it suffices to compute it in one case.
Ignoring finite size effects\footnote{In principle, the measurements near the ends of the repeater line have smaller error probabilities. Ignoring this slightly underestimates performance, however, the difference is vanishingly small for a large number of repeater stations.} and potential correlations between the error probabilities of different repeater stations, we estimate the final $X$-error distribution $\mathcal{P}_\text{fin}(X) =  \mathcal{P}^{\ast N}_\text{sq}(X, \sigma^2) $ on Bob's qudit as the $N$-fold discrete convolution of $\mathcal{P}_\text{sq}(X, \sigma^2)$, where $N$ denotes the number of repeater stations.
We expect that this estimate captures the general behavior of the performance of GKP qudit repeaters.
In principle, computing this convolution can be sped up by diagonalizing the corresponding error-probability matrix~\cite{MHKB18}.
For our purposes, however, a direct implementation is sufficient.
Then, we compute the outer product $\mathcal{P}_\text{fin}(X, Z) = \mathcal{P}_\text{fin}(X) \otimes \mathcal{P}_\text{fin}(Z)$.
The secret-key rate of the repeater line, finally, is given by $\log_2(D)-H(\mathcal{P}_\text{fin}(X, Z)) = \log_2(D)-2 H(\mathcal{P}_\text{fin}(X))$.

\section{Error analysis of GKP repeaters with higher-level codes} 
\label{app:error_analysis_encoded}

In this appendix, we lift our error analysis from App.~\ref{app:error_analysis_bare}
to the logical level.
First, we discuss in App.~\ref{app:teleportation_encoded_analysis} the two repeater protocols from Sec.~\ref{sec:2way_postamp} and~\ref{sec:1way_preamp}.
In App.~\ref{app:half_teleportation_gkp_syndrome_placement}, 
we discuss the optimal placement of the lower-level GKP measurements for the third protocol from Sec.~\ref{sec:1way_preamp_half} and analyze its performance.

\subsection{Logical performance of GKP qudits concatenated with quantum polynomial codes}
\label{app:teleportation_encoded_analysis}
In App.~\ref{app:error_analysis_bare}, 
we showed that the error probability distribution for measurements in repeater stations is given by $\mathcal{P}_\text{sq}(X, \sigma^2)$, where $\sigma_\text{2-way}^2 = 3\sigma_\text{sq}^2+\frac{1}{\eta_\text{c} \sqrt{\eta}}-1$ and
$\sigma_\text{1-way}^2 = 3\sigma_\text{sq}^2 + 1-\eta_\text{c}\eta$ 
for the two-way and one-way teleportation protocol, respectively.
When the protocol is lifted to its logical version,
we still find the same error distribution for each of the measurements of the physical GKP qudits (of which there are $D$).
This is because $\overline{CZ} = (CZ^\dagger)^{\otimes D}$ is semitransversal for the quantum polynomial code with parameters $\llbracket D, 1,\tfrac{D+1}{2} \rrbracket_D$~\cite{Aharonov2008}.

Here, we consider a simple decoder that only corrects errors occurring on a number of qudits not more than half the distance $d=\tfrac{D+1}{2}$.
Thus, the probability that a correctable error pattern occurs at a repeater station is given by 
\begin{align} \label{eq:correctable_error_probability}
    p_\text{cor} = \sum_{k=0}^{\tfrac{d-1}{2}} \binom{D}{k} p_0^{D-k}(1-p_0)^k,
\end{align}
where $p_0 = P_\text{sq}(X^0, \sigma^2)$, as in Eq.~\eqref{eq:pauli_shift_probability_gauss}.
If the decoding attempt fails, we replace the measured state with the maximally mixed state (as a worst-case approximation).
In other words: with probability $1 - p_\text{cor}$, we insert a logical error, uniformly at random from the set $\{1,\ldots, D-1\}$.
Therefore, the error probability distribution on measurement outcomes in any repeater station is given by 
\begin{align}\label{eq:repeater_error_channel}
    P_\text{rep} (X^k) = \begin{cases} p_\text{cor} & \text { if }k=0 \\
    \frac{1}{D-1}(1-p_\text{cor}) & \text { otherwise}.
    \end{cases}
\end{align} 
If the probability of errors is so large that $p_\text{cor}<\frac{1}{D-1}(1-p_\text{cor})$, 
we replace Eq.~\eqref{eq:repeater_error_channel} by the uniform distribution.
Again, ignoring correlations between error distributions on different repeater stations,
we estimate the final error distribution of the encoded repeater line as $\mathcal{P}_\text{fin}(X, Z) = \mathcal{P}^{\ast N} _\text{rep}(X) \otimes \mathcal{P}^{\ast N} _\text{rep}(Z)$.
% and the secret-key rate of the quantum repeater line as $\log_2(D)-2 H(\mathcal{P}^{\ast N}_\text{rep}(X))$.

\subsection{Error analysis of the half-teleportation protocol for various placements of GKP syndrome measurements}
\label{app:half_teleportation_gkp_syndrome_placement}

In this appendix, we discuss how introducing additional ancilla-based measurements of lower-level GKP stabilizers can improve the performance of the one-way half-teleportation protocol with optical pre-amplification from Sec.~\ref{sec:1way_preamp_half}.
Such measurements are pictured in Fig.~\ref{fig:repeaters}~(c) of the main text.
As discussed in Sec.~\ref{sec:loss_and_amplification}, 
every transmission from one repeater station to the next is associated with a Gaussian error channel with variance $\sigma^2_\text{loss} = 1- \eta_\text{c}\eta$, where $\eta = \exp(-L_0/L_\text{att})$.
In every repeater station, all incoming GKP qudits are measured in the $p$-quadrature.
Before this, however, each GKP qudit is coupled via a physical $\CPhase^{\dagger}$-gate to a qudit in the next logical block.
Since the $\CPhase^{\dagger}$-gate spreads $p$-errors into $q$-errors, but $q$-errors are not propagated to the next mode, every error source only has a limited range.
A $p$-error that arises during one transmission, does not directly affect $p$-measurements on the qudit it occurred to, however, it propagates into a $q$-error on the subsequent GKP qudit, which alters the $p$-measurement outcome of that qudit.
Furthermore, 
a $p$-error during GKP state preparation backpropagates through the $\CPhase^{\dagger}$-gate and causes a $q$-error on the readout of the preceding GKP qudit.

\begin{figure}[t]
    \centering	
    \includegraphics[scale=.9]{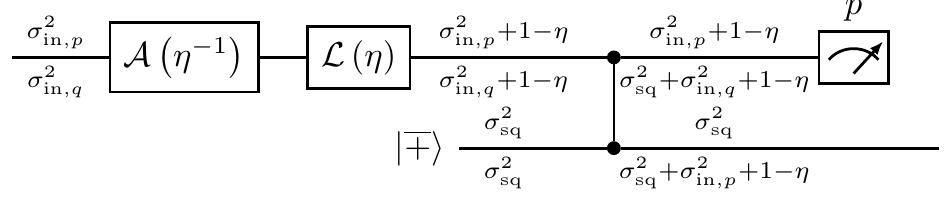}
    \caption{Propagation of Gaussian errors for the half-teleportation protocol \emph{without} additional GKP stabilizer measurements.
    Because of periodic boundary conditions, we have 
    $\sigma^2_{\text{in},p} = \sigma^2_{\text{out},p} = \sigma^2_\text{sq}$ and 
    $\sigma^2_{\text{in},q} = \sigma^2_{\text{out},q} = 2\sigma^2_\text{sq} + 1-\eta$.
     Therefore, the variance of $q$-errors reaching the $p$-measurements is given by $\sigma^2_\text{sq}+\sigma^2_{\text{in},q} +1-\eta = 3\sigma^2_\text{sq} + 2(1-\eta)$. }
    \label{fig:error_analysis_no_extra}
\end{figure}

\begin{figure}[t]
    \centering	
    \includegraphics[scale=.9]{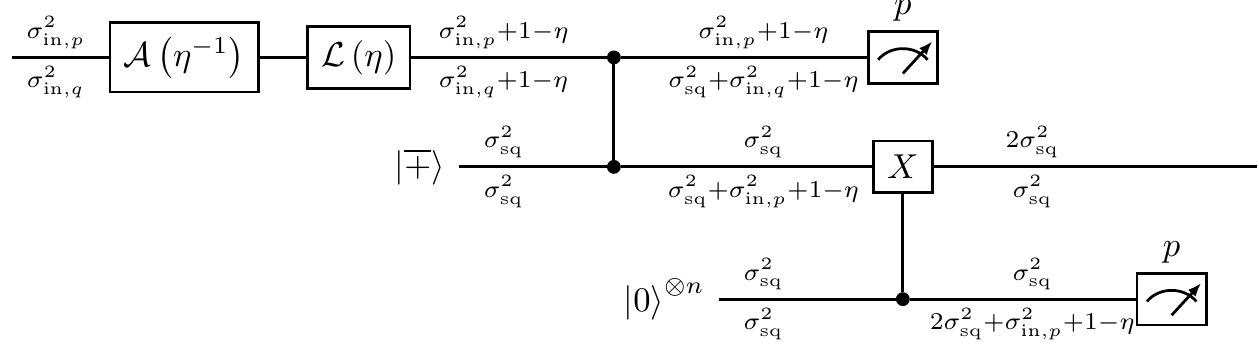}
    \caption{Propagation of Gaussian errors for the half-teleportation protocol with additional GKP stabilizer measurements \emph{after} every $CZ$-gate.
    Because of periodic boundary conditions, we have 
    $\sigma^2_{\text{in},p} = \sigma^2_{\text{out},p} = 2\sigma^2_\text{sq}$ and 
    $\sigma^2_{\text{in},q} = \sigma^2_{\text{out},q} = \sigma^2_\text{sq}$.
     Therefore, the variance of $q$-errors reaching the $p$-measurements is given by 
     $\sigma^2_\text{sq}+\sigma^2_{\text{in},q} + 1-\eta = 2\sigma^2_\text{sq} + 1-\eta$.
     In addition to these continuous displacement errors, a discrete Pauli error channel $\mathcal{P}_\text{sq}(Z, \sigma^2_\text{GKP})$ leads to lower-level logical errors on every $X$-measurement, where $\sigma^2_\text{GKP} = 2\sigma^2_\text{sq} + \sigma^2_{\text{in},p} + 1-\eta = 4\sigma^2_\text{sq}+1-\eta$ is the variance of $q$-errors reaching the lower-level GKP stabilizer measurement.
     }
    \label{fig:error_analysis_always_after}
\end{figure}

\begin{figure}[t]
    \centering	
    \includegraphics[scale=.9]{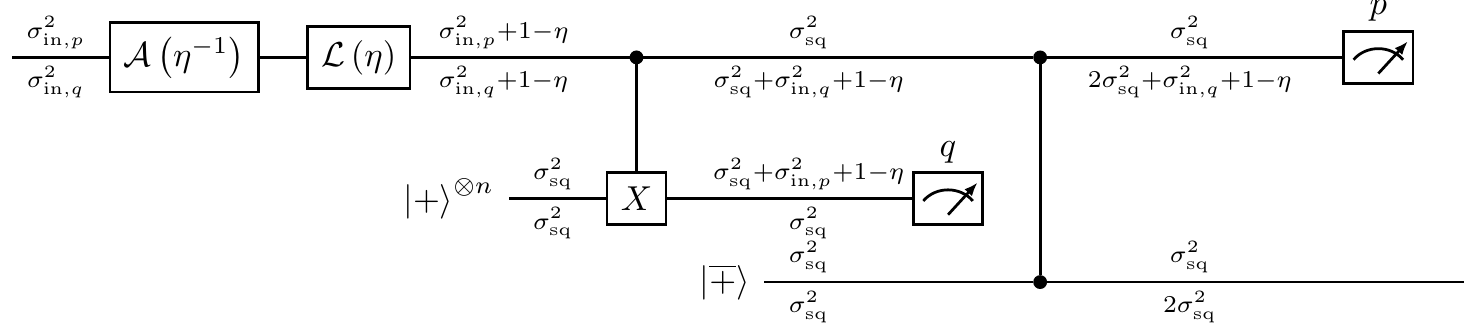}
    \caption{Propagation of Gaussian errors for the half-teleportation protocol with additional GKP stabilizer measurements \emph{before} every $CZ$-gate.
    Because of periodic boundary conditions, we have 
    $\sigma^2_{\text{in},p} = \sigma^2_{\text{out}, p} = \sigma^2_\text{sq}$ and 
    $\sigma^2_{\text{in},q} = \sigma^2_{\text{out}, q} = 2\sigma^2_\text{sq}$.
     Therefore, the variance of $q$-errors reaching the $p$-measurements is given by 
     $2\sigma^2_\text{sq}+\sigma^2_{\text{in},q} + 1-\eta = 4\sigma^2_\text{sq} + 1-\eta$.
     In addition to these continuous displacement errors, a discrete Pauli error channel $\mathcal{P}_\text{sq}(Z, \sigma^2_\text{GKP})$ leads to lower-level logical errors on every $X$-measurement, where $\sigma^2_\text{GKP} = \sigma^2_\text{sq} + \sigma^2_{\text{in},p} + 1-\eta = 2\sigma^2_\text{sq}+1-\eta$ is the variance of $p$-errors reaching the lower-level GKP stabilizer measurement. Originally, the lower-level GKP stabilizer measurement results in a discrete Pauli error channel $\mathcal{P}_\text{sq}(X, \sigma^2_\text{GKP})$, which is then propagated to a Pauli error channel $\mathcal{P}_\text{sq}(Z, \sigma^2_\text{GKP})$ in the next segment due to the $CZ$-gate.}
    \label{fig:error_analysis_always_before}
\end{figure}

\begin{figure}[t]
    \centering	
    \includegraphics[width=\textwidth]{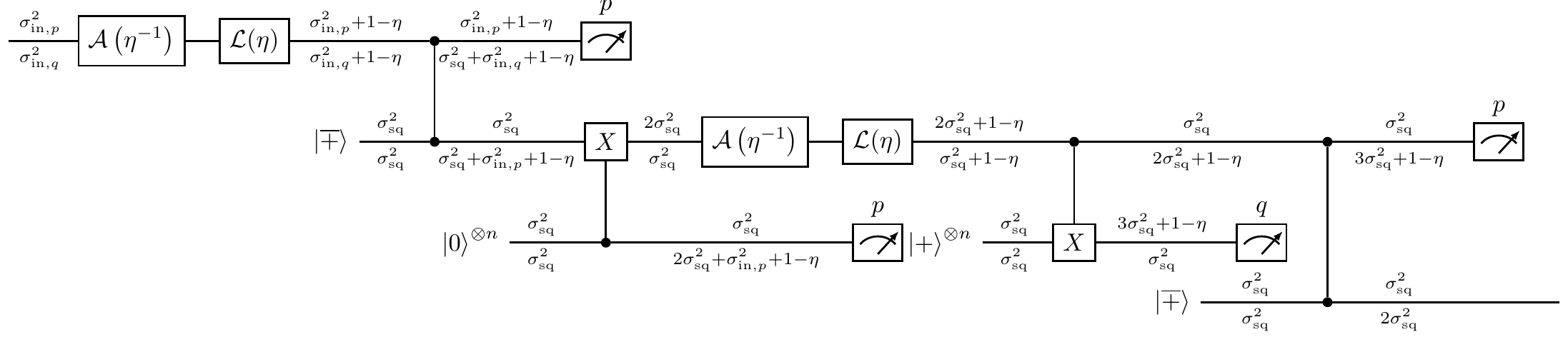}
    \caption{Propagation of Gaussian errors for the half-teleportation protocol with additional GKP stabilizer measurements at \emph{alternating} placements.
    Because of periodic boundary conditions, we have 
    $\sigma^2_{\text{in},p} = \sigma^2_{\text{out}, p} = \sigma^2_\text{sq}$ and 
    $\sigma^2_{\text{in},q} = \sigma^2_{\text{out}, q} = 2\sigma^2_\text{sq}$.
    Therefore, it turns out that the variance of $q$-errors reaching all $p$-measurements is given by 
     $\sigma^2_\text{sq}+\sigma^2_{\text{in},q} + 1-\eta = 3\sigma^2_\text{sq} + 1-\eta$.
   In addition to these continuous displacement errors, a discrete Pauli error channel $\mathcal{P}_\text{sq}(Z, \sigma^2_\text{GKP})$ leads to lower-level logical errors on every $X$-measurement, where $\sigma^2_\text{GKP} = 2\sigma^2_\text{sq} + \sigma^2_{\text{in},p} + 1-\eta 
   = 3\sigma^2_\text{sq}+1-\eta$ is the variance of errors reaching and altering lower-level GKP stabilizer measurements.
    }
    \label{fig:error_analysis_alternating}
\end{figure}

In the plain version (without lower-level GKP stabilizer measurements),
errors on physical readouts (in the repeater stations) follow an error distribution $\mathcal{P}_\text{sq}(Z,  2\sigma^2_\text{loss}+3\sigma^2_\text{sq})$, where the variance takes noise from two transmissions and three GKP state preparations into account.
By introducing a lower-level GKP stabilizer measurement in every repeater station, we can correct displacement errors after a single transmission.
In this way, we effectively avoid combining the two transmission loss channels.
Instead, all Gaussian errors in one quadrature are replaced by the discrete Pauli error channel from Eq.~\eqref{eq:pauli_shift_probability_gauss}.
Such discrete qudit Pauli errors will propagate to the measurements in the usual way~\cite{MHKB18}.
Depending on where in the repeater station we place the ancilla-based GKP stabilizer measurement, the final error distribution will vary.
We discuss four options:
\begin{enumerate}[label=(\roman*)]
    \item No additional GKP stabilizer measurements are performed, see Fig.~\ref{fig:error_analysis_no_extra} for the error analysis.
    \item \emph{After} every $CZ$-gate, the (physical) target qudit is subjected to a GKP stabilizer measurement of $S_X= \exp(-\text{i}\sqrt{2\pi D} \hat p)$.
    This is achieved by preparing an ancillary GKP qudit in state $\ket{0}$, applying a $\CSUM$-gate from the ancilla to the repeater qudit, and a $p$-measurement of the ancilla GKP qudit, see Fig.~\ref{fig:error_analysis_always_after} for the error analysis.
    
    \item  \emph{Before} every $CZ$-gate, the control qudit is subjected to a GKP stabilizer measurement of $S_Z= \exp(\text{i}\sqrt{2\pi D} \hat q)$.
    This is achieved by preparing a  GKP ancilla in state $\ket{+}$, applying a $\CSUM$-gate from the repeater qudit to the ancilla, followed by a $q$-measurement of the ancilla, see Fig.~\ref{fig:error_analysis_always_before} for the error analysis.
    \item We \emph{alternate} between options (ii) and (iii), see Fig.~\ref{fig:error_analysis_alternating} for the error analysis.
\end{enumerate}
In option~(i), the error analysis from App.~\ref{app:teleportation_encoded_analysis} with $\sigma^2 =  2\sigma^2_\text{loss}+3\sigma^2_\text{sq}$ applies, see Fig.~\ref{fig:error_analysis_no_extra}.
Both in option~(ii) and~(iii), which we refer to as \emph{symmetric} placements of the GKP stabilizer measurements, it turns out that every $p$-measurement is subject to two discrete Pauli error channels as in Eq.~\eqref{eq:pauli_shift_probability_gauss}, one having variance $2\sigma^2_\text{sq}+\sigma^2_\text{loss}$ and the other one $4\sigma^2_\text{sq}+\sigma^2_\text{loss}$.
Thus, the error analysis from App.~\ref{app:teleportation_encoded_analysis} applies after we insert 
\begin{align} \label{eq:symmetric_placement_error_probs}
p_0^\text{sym} = \sum_{k=0}^{D-1} P_\text{sq}(X^k, 2\sigma^2_\text{sq}+\sigma^2_\text{loss}) P_\text{sq}(X^{D-k}, 2\sigma^2_\text{sq}+\sigma^2_\text{loss})
\end{align}
into Eq.~\eqref{eq:correctable_error_probability}.
Finally, in option~(iv) both GKP stabilizer and logical measurements are subject to Gaussian errors with variance  $3\sigma^2_\text{sq}+\sigma^2_\text{loss}$.
This time, we thus have to insert 
\begin{align} \label{eq:alternating_placement_error_probs}
p_0^\text{alt} = \sum_{k=0}^{D-1} P_\text{sq}(X^k, 3\sigma^2_\text{sq}+\sigma^2_\text{loss}) P_\text{sq}(X^{D-k}, 3\sigma^2_\text{sq}+\sigma^2_\text{loss})
\end{align}
into Eq.~\eqref{eq:correctable_error_probability}.

% In option~(i), the error analysis from App.~\ref{app:teleportation_encoded_analysis} with $\sigma^2 =  2\sigma^2_\text{loss}+3\sigma^2_\text{sq}$ applies, see Fig.~\ref{fig:error_analysis_no_extra}.
% In option~(ii), an ancilla-based measurement of $S_X= \exp(-\text{i}\sqrt{2\pi D} \hat p)$ is performed after each $CZ$-gate.
% In Fig.~\ref{fig:error_analysis_always_after}, we show how displacement errors propagate to the $p$-measurement in any given repeater station.
% By denoting the variances of initial displacement errors by $\sigma^2_{\text{in},q}$ and  $\sigma^2_{\text{in},p}$, we find that the final $p$-variance affecting the $p$-measurement is given by $\sigma^2_{\text{in},p}+2\sigma^2_\text{sq} +1-\eta$.
% Furthermore, the variances at the end of the circuit are given by $\sigma^2_{\text{out},q}=\sigma^2_\text{sq}$ and $\sigma^2_{\text{out},p} =  2\sigma^2_\text{sq}$.
% Since periodic boundary conditions apply, i.e., $\sigma^2_{\text{out},q} = \sigma^2_{\text{in},q} $ and $\sigma^2_{\text{out},p} = \sigma^2_{\text{in},p} $, 
% the continuous displacement errors arriving at the $p$-measurements have a variance of $4 \sigma^2_\text{sq} +1 - \eta$.

% we introduce a Pauli error channel where $Z$-errors are distributed according to $\mathcal{P}(Z,\sigma^2_\text{loss}+4\sigma_\text{sq}^2)$...

% % Show Figures for propagation of variances

\begin{figure} \centering
	\subfloat[]{\label{fig:skr_VS_L_with_L=2000km_20_squeezing0.999_coupling_noerasure_twoway}\includegraphics[height=0.275\textwidth]{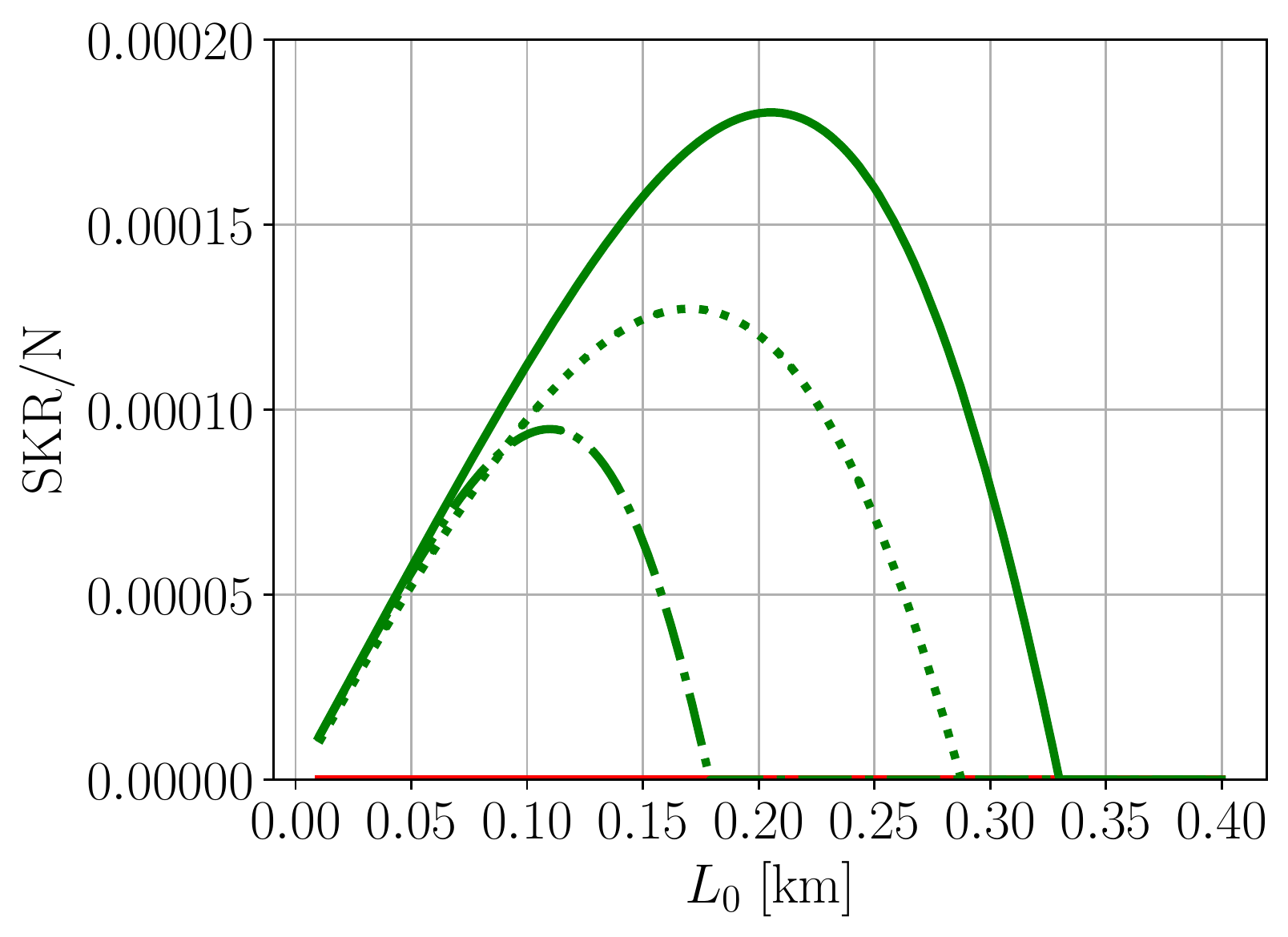}}
    \hspace{5mm}
	\subfloat[]{\label{fig:skr_VS_L_with_L=2000km_30_squeezing0.999_coupling_noerasure_twoway_app}\includegraphics[height=0.275\textwidth]{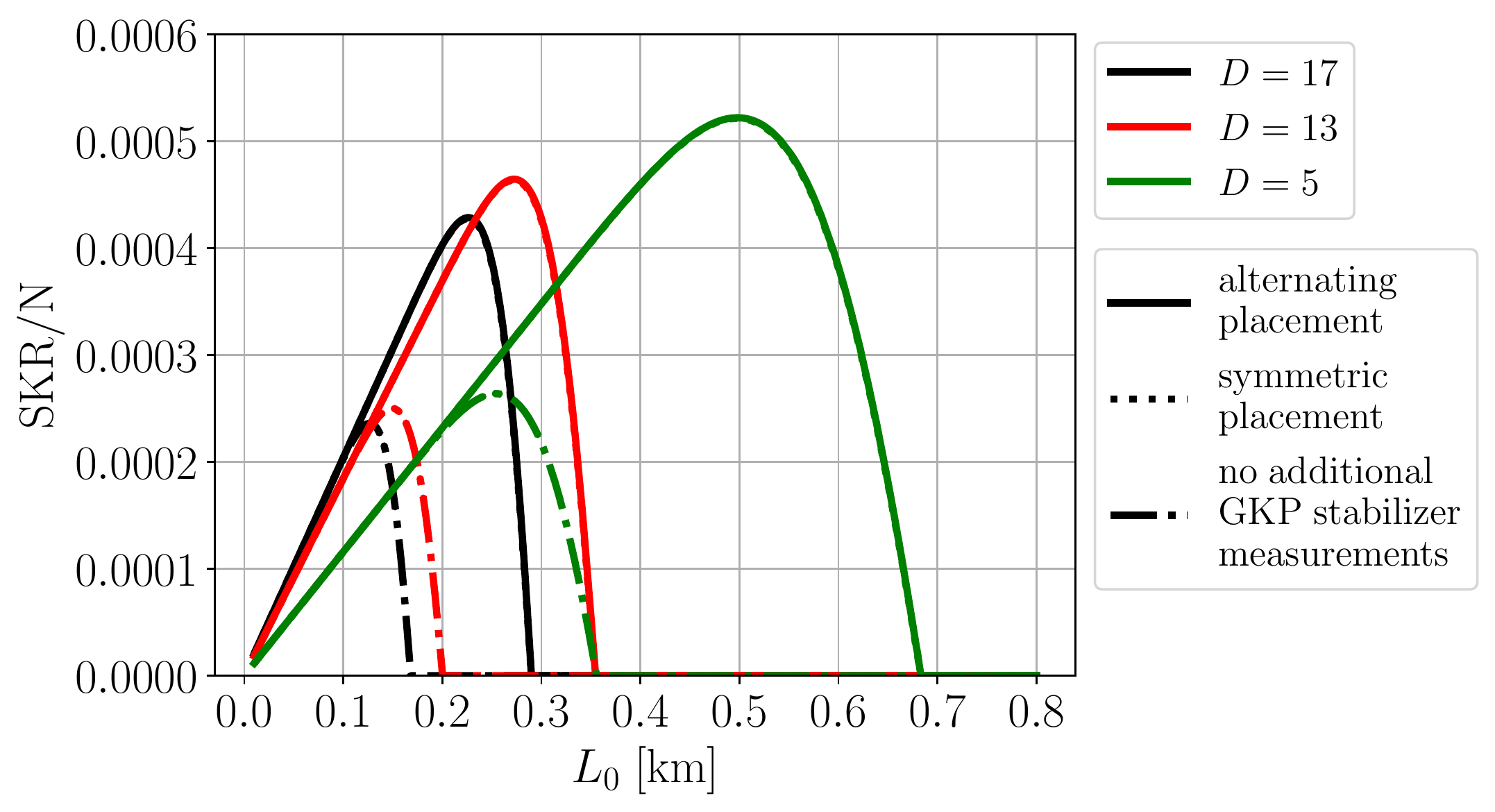}}\\
	\caption{Lower bound on the SKR per logical channel use, $\log_2(D)-H(\mathcal{P})$, normalized by the number $N$ of repeater stations for the one-way half-teleportation protocol and various placements of lower-lever GKP stabilizer measurements.
    We plot SKR$/N$ as a function of the repeater spacing $L_0$ for a repeater line of total length $L = N  L_0 = 2000\unitspace\text{km}$, coupling efficiencies $\eta_\text{c}=99.9\%$, and squeezing levels of (a) $s_\text{GKP} = 20\unitspace\text{dB}$ or (b) $s_\text{GKP}=30\unitspace\text{dB}$.
	}
	\label{fig:figappendix}
\end{figure}

In Fig.~\ref{fig:figappendix}, we show how the placement of GKP stabilizer measurements influences the performance of the half-teleportation protocol, using the exact same setting as in Fig.~\ref{fig:figc} of the main text.
Overall, the situation is very similar to that in Fig.~\ref{fig:figc}: for $s_\text{GKP}=20\unitspace\text{dB}$ in Fig.~\ref{fig:figappendix}~(a), only the $\llbracket5,1,3\rrbracket_D$-code (green) offers a nonzero SKR, whereas  for $s_\text{GKP}=20\unitspace\text{dB}$ in Fig.~\ref{fig:figappendix}~(b) also the $\llbracket13,1,7\rrbracket_D$-code (red) and the  $\llbracket17,1,9\rrbracket_D$-code (black) have the potential to distribute secret keys.
We see in Fig.~\ref{fig:figappendix} that  an alternating placement of GKP stabilizer measurements (solid curves) leads to the highest values of SKR$/N$.
For both option~(ii) and~(iii), the symmetric placements (dotted curves) are governed by Eq.~\eqref{eq:symmetric_placement_error_probs}, and therefore lead to the same performance.
We see that not performing any additional GKP stabilizer measurements (dash-dotted curve) 
leads to the lowest performance, which is easily explained by the large variance $2\sigma^2_\text{loss}+3\sigma^2_\text{sq}$.
The other options break the term $2\sigma^2_\text{loss}$ and, therefore, perform better.
For the symmetric placement, the bottleneck is posed by the term $4\sigma^2_\text{sq}$ in Eq.~\eqref{eq:symmetric_placement_error_probs}, which is worse than 
$3\sigma^2_\text{sq}$ in Eq.~\eqref{eq:alternating_placement_error_probs} for the alternating placement.
This explains why the latter performs best.
For a large squeezing value of $s_\text{GKP}=30\unitspace\text{dB}$, the difference between $3\sigma^2_\text{sq}$ and $4\sigma^2_\text{sq}$ is negligible, which causes a nearly perfect overlapping of the dotted and solid curves in Fig.~\ref{fig:figappendix}~(b).

Since the alternating placement of GKP stabilizer measurements has the best performance, we have assumed this option for the one-way half-teleportation protocol throughout the main text of this paper.

\section{Author contributions}

DM initiated the project as a whole and exploring the idea of concatenating GKP qudits with polynomial codes.
FS designed the repeater protocols, derived the analytical model, performed the numerics, and created the figures.
FS and DM designed the study, interpreted the results, and wrote the manuscript. 
PvL supported research and development and helped preparing
the manuscript.

%\appendix  
   
\newpage
\bibliography{ref}   
\bibliographystyle{quantum}

\end{document}